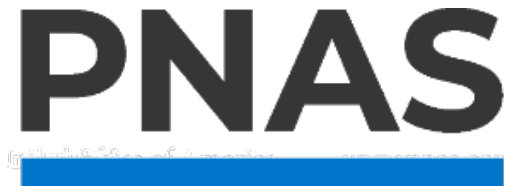

# Mercury's Crustal Magnetization Indicates a Stronger Ancient Dynamo

Isaac S. Narrett[1]*, Benjamin P. Weiss[1], Sarah C. Steele[2], John B. Biersteker[1]

[1]Department of Earth, Atmospheric, and Planetary Sciences, Massachusetts Institute of Technology, Cambridge, MA, USA.
[2]Department of Earth and Planetary Sciences, Harvard University, Cambridge, MA 02138, USA.
*Corresponding author. Email: Isaac S. Narrett, narrett@mit.edu

**Author contributions:**

Conceptualization: ISN, BPW, SCS, JBB, Methodology: ISN, BPW, SCS, JBB, Investigation: ISN, BPW, SCS, JBB, Visualization: ISN, Funding acquisition: ISN, BPW, Project administration: BPW, JBB, Supervision: BPW, JBB, Writing – original draft: ISN, Writing – review & editing: ISN, BPW, SCS, JBB, Data Curation: – ISN, SCS, Validation: – ISN, SCS, Software: – ISN, SCS, JBB, Formal Analysis: – ISN, BPW, SCS, JBB, Resources: – ISN, BPW

**Competing Interest Statement:** The authors declare no competing interests.

**This file includes:**

Main Text
Figures 1 to 6

**Abstract**
Mercury is the only terrestrial planet in the solar system other than Earth with an active dynamo magnetic field (~200 nT at the equatorial surface). Furthermore, Mercury's ~3.9-3.7-billion-year-old (Ga) crust is strongly magnetized (~10 nT at ~30-km altitude), indicating the presence of a past dynamo. However, the strength of the past dynamo field and the mechanism that generated it are unknown. To address this, we performed three-dimensional magnetohydrodynamic simulations of the ancient solar wind interaction with the planetary field coupled with crustal thermal evolution and magnetization models. We show that the crustal magnetization was likely produced by a dipole field with equatorial surface strength of at least ~2,000 nT and possibly as high as ~30,000 nT for a dynamo with a reversal frequency greater than once per million years. Such strong fields likely exclude both the solar wind feedback and thermoelectric dynamo mechanisms at 3.7 Ga ago. Instead, our results are compatible with the past dynamo being generated by a nearly fully convective core.

**Significance Statement**

Mercury's present-day dynamo-generated magnetic field is anomalously weak. However, Mercury's 3.9-3.7-billion-year-old crustal remanent magnetization suggests that the planet once generated a dynamo magnetic field that was ~10-100 times stronger than its field today. The strength of this ancient, planet-scale magnetic field implies that the dynamo was once driven by an Earth-like mechanism involving strong core convection. Subsequently, the driving energy weakened to produce the present-day, weak field.

**MAIN TEXT**

**Introduction**

Magnetic field measurements by the Mariner 10 (1) and MErcury Surface, Space ENvironment, GEochemistry, and Ranging [MESSENGER (2)] spacecraft indicate that Mercury has a planetary-scale magnetic field generated by a core dynamo while its crust contains natural remanent magnetization (NRM) (3, 4). The dynamo field is approximated as a northward planetocentric-offset, spin-axis aligned, dipole with a field of ~200 nT at the magnetic equatorial surface (3). Given the available energy and core size (5-7), this field is ~2 orders of magnitude weaker than that expected from core convection scaling laws [e.g., Elsasser number ~ 1 convection scaling based on dynamo Lorentz and Coriolis force balance (8, 9)]. Motivated by these peculiarities, a diversity of dynamo mechanisms [e.g., thermoelectric (10, 11), thin-shell convection (12, 13), thick-shell convection (14), convection under a thick stable layer with uniform heat flux (15-17), iron-snow (18-20), double-diffusive convection (17, 21, 22), solar wind feedback (23-25), and convection (26) and/or stable layer (27) with variable heat flux] have been proposed to explain the present-day magnetic field [Table S1, also see section 3.5 of (28) for a recent extensive review]. However, the nature and origin of the ancient field are currently largely unknown.

Studying Hermean crustal magnetization can provide insights into the ancient state of the dynamo and interior. Previous studies (4, 29) have implicitly showed that a stronger ancient dynamo could have magnetized the Hermean crust, but they could not differentiate between an Earth-strength dynamo field (~50,000 nT) and the present-day Mercury dynamo field (~200 nT) as the magnetizing source. This motivates the goal of this study: to quantify the magnitude of the ancient paleofield and explore its implications for Mercury's dynamo and interior evolution.

MESSENGER's low-altitude (<100-km) passes identified strong crustal fields [peak strength ~10 nT at ~30-km; see Figures 5.16 and 5.19 in (4)] located at latitudes >35° N, with insufficient low altitude data southward of ~35° N. The crustal anomalies are generated from NRM associated with near-surface impact structures and volcanic flows (29-34). Although shock remanent magnetization (SRM) may be present in the Hermean surface, SRM alone, even in combination with a dynamo amplified by impact-produced plasmas, is unlikely to explain all of the widespread, strong, and geologically heterogeneous crustal fields observed in the northern hemisphere (35) (see Fig. S1). In line with past interpretations, the ancient (3.9–3.7 Ga old) northern hemisphere volcanic smooth plains (36-38) likely carry thermoremanent magnetization (TRM), acquired as tens of km wide, <70-km thick (34) volcanic structures cooled over millions of years (Ma). For TRM in these large volumes, the magnetizing paleofield is typically assumed to have been present and stationary (or at least with a low reversal rate) over such a slow cooling time (39), an attribute commonly ascribed to the existence of an ancient (paleo) dynamo.

Alternatively, the ancient interplanetary magnetic field (IMF) is also a candidate magnetizing field given that it should have been similar in strength (300-600 nT) to the present-day surface dipole field (200-800 nT) (see Materials and Methods). Although the impact plasma-amplified IMF could be recorded as an SRM in Mercury's crust (35), the IMF could potentially have produced much stronger crustal TRM given the >10-100 times higher recording efficiency of TRM relative to SRM (40). If the IMF was the field that magnetized the Hermean crust, this would challenge the assumption that a core dynamo was active when Mercury's crustal TRM was acquired. This in turn would imply that the crustal TRM is not a record of the planet's interior thermal and structure evolution. The possibility that the IMF and/or a dynamo contributed to Mercury's past field can be assessed by quantifying the field's paleointensity and self-consistently modeling the TRM recording process. Furthermore, as discussed below, because different hypothesized dynamo mechanisms predict different ancient field strengths, constraining the paleofield intensity may help differentiate between the mechanisms that powered Mercury's past dynamo.

In the absence of definitive Mercury meteorite samples for paleomagnetic study (41), the only available rock record is the crustal TRM in Mercury's northern hemisphere. Until now, due to the non-uniqueness associated with inverting magnetic field data for magnetization, both the crustal magnetic mineralogy (which determines the TRM recording efficiency, $\chi_{TRM}$) and the paleofield strength ($B_{paleo}$) remain poorly constrained, spanning $\chi_{TRM}$ ~ $10^{-4}$ – 1 and $B_{paleo}$ ~ $10^{1}$ – $10^{5}$ nT, respectively (4, 42). As we will show, such a large uncertainty in the paleofield intensity does not meaningfully distinguish between any of the dynamo mechanisms and the IMF.

Previous work (29) derived the aforementioned range for $B_{paleo}$ by estimating an extreme lower bound using an idealized mathematical framework. Specifically, that framework assumes (i) a constant magnetization magnitude with (ii) idealized geometry in which the magnetization direction varies spatially to maximize the field at the spacecraft measurement location and (iii) a laterally infinite source layer with finite thickness (43). These assumptions are imposed to minimize the magnetization required to generate the observed crustal anomaly, but are not physically justified to explain TRM recorded in the ancient Hermean planetary field environment. In particular, the extreme solar wind forcing and the likelihood of dynamo reversals would cause the intensity and geometry of the magnetospheric field (i.e., the superposition of the planetary dipole field, induced field in the planetary interior, magnetospheric current generated fields, and other plasma-generated field sources) to vary in the source region as the TRM is acquired, violating assumptions (i) and (ii), respectively [e.g., see Figure 1 in (43)]. Similarly, because crustal fields of comparable magnitude are observed at multiple locations (Fig. S1), assumption (ii) would require the TRM to be configured as to maximize the field at each observation point, which is not a physically plausible outcome of TRM acquisition in a planetary dynamo field. Moreover, crustal TRM on Mercury is finite in lateral extent and requires finite magnetized volume calculations, thus violating assumption (iii).

To apply a more physically-motivated framework to explain these measurements, we quantify $B_{paleo}$ by self-consistently modeling the magnetization process on Mercury. We achieve this by translating recent laboratory experiments into constraints on Hermean magnetic mineralogy ($\chi_{TRM}$) and performing first-of-their-kind, coupled simulations that track the time-evolution of ancient Mercury's surface magnetic field environment and magnetization. This allows us to distinguish between the sources of the Hermean paleomagnetic field, to differentiate between the proposed dynamo mechanisms, and to understand the thermal and interior evolution over time.

**Mercury's Magnetic Mineralogy.** To model the crustal magnetization, it is essential to evaluate the possible magnetic mineralogy of the crust and its TRM recording properties ($\chi_{TRM}$). Measurements of the average crustal Fe content [~1.5 wt% (44-46)] and compositional features (e.g., sulfide speciation) indicate that Mercury formed and differentiated under extremely reducing conditions [with an oxygen fugacity, log $fO_2$, that was 3 to 7 log units below that of the iron-wüstite (IW) buffer (47)]. Although MESSENGER measured a relatively high sulfur content [~1.5-4 wt% (45)], laboratory experiments and modeling of sulfide speciation under Mercury's reducing conditions predict that only a small fraction (<<5%) of the available sulfur will be in iron sulfide [and even this is likely only found predominantly in regions where log $fO_2$ ~IW - 3 or higher; see Fig. 7 of (48)]. Due to the extremely reducing environment, most of the available sulfur should have formed non-iron sulfides like $Na_2S$, CaS, MgS, MnS, and NaCrS (48-51). Furthermore, the ~593 K Curie temperature of monoclinic ($Fe_7S_8$) pyrrhotite (a ferromagnetic iron sulfide) would only allow TRM to be held in layers shallower than ~10-km at latitudes ≲60° N and S at >3 Ga ago given the average orbital configuration [Figure 5 in (42)]. As such, monoclinic pyrrhotite as well as the iron sulfide greigite [Curie temperature ~677 K (52)] are likely not the primary magnetic carriers in Mercury's crust [although greigite has been demonstrated to be thermodynamically stable in Mercury's interior (53)]. Some magnetic carriers like daubréelite ($FeCr_2S_4$) and cohenite ($Fe_3C$) are also likely not dominant contributors to surface magnetization due to their relatively low Curie temperatures (~150 K and ~483 K, respectively), such that they could only hold a stable TRM in ≲1-km layers [too shallow based on some measured crustal signals (4, 34); see Figure 3 in (42) for Curie depth calculations]. Other magnetic carriers like schreibersite ($Fe_3P$) and suessite ($Fe_3Si$) could exist on Mercury, but their TRM recording properties remain poorly understood (42).

In contrast, Fe-metal and FeNi-metal phases (e.g., kamacite and martensite), have sufficiently high Curie temperatures (~1043 K) to retain a remanence through most or all of the ~70-km thick Hermean crust (54, 55). For probable current and ancient orbital configurations [eccentricity ~0-0.4 (56-58)] and geotherms (4-10 K $km^{-1}$), the Fe-metal Curie depth would be ≥70-km [see Figures 2 and 3 in (42) for Curie depth calculations]. This Fe mineralogy is also known to dominate the magnetic recording carriers of the aubrite meteorites (59, 60), which are inferred to form in reducing conditions similar to that of Mercury [~IW-4 (51)] while possessing a similar to slightly higher bulk iron content [~1-10 wt% (61)] compared to Mercury's crust [~1.5 wt% Fe (44)]. Therefore, we consider aubrites as a reasonable magnetic analog to Mercury's crustal material, with the Fe- and FeNi-metal phase lithologies capable of recording and retaining a crustal TRM.

To estimate the TRM recording efficiency for the aubrite-like composition, we relate measurements of saturation remanent magnetization, $M_{RS}$, to $\chi_{TRM}$, and find $\chi_{TRM}$ ~ 0.0003-0.02 [(60, 62-65); see Supporting Information Text for details on these calculations]. We report our subsequent crustal field calculations using the upper limit of the aubrites TRM recording efficiency ($\chi_{TRM}$ ~ 0.02), and consequently, our inferred $B_{paleo}$ values are likely underestimated. Additionally, because crustal field strength is proportional to $\chi_{TRM}$ [equation 1 in the supplementary material of (64)], the crustal field calculations can be scaled to different materials with their respective $\chi_{TRM}$.

It is possible that some localized regions (~10-100-km) of the Hermean surface could be enriched in Fe-metal with respect to the global average and have a correspondingly higher $\chi_{TRM}$ ~ $10^{-1}$ [e.g., from impactor-delivered iron (30-33)]. Such localized Fe-metal enrichment from impactors has been proposed to have produced the sparse, strong crustal anomalies associated with impact structures on the Moon (66, 67), as endogenous lunar materials are comparatively poor magnetic recorders [$\chi_{TRM} \leq 10^{-3}$ (64)]. However, unlike the Moon, MESSENGER measurements suggest Mercury's magnetic anomalies are widespread and co-located with varying geological structures [e.g., volcanic smooth plains, areas with low density of impact craters; see Fig. S1 and Fig. 5.19 from (4)]. Utilizing $\chi_{TRM}$ ~ 0.02 (see above) for modeling the magnetization process only requires that at least one TRM unit formed with global-average Fe-metal abundance, which is highly likely given the widespread occurrence of crustal anomalies and the lack of spatial association between them, their geologic setting, and crustal Fe-metal abundance [(44, 46); see also Fig. S1].

**Mercury's Magnetic Field Environment.** A second important consideration for modeling the ancient magnetization is the surface magnetic field environment ($B_{surface}$) around 3.9-3.7 Ga ago. The young, active Sun had both a higher mass loss rate (MLR) and higher average surface magnetic field at this time, meaning that the emitted solar wind plasma was more energetic and carried a stronger IMF compared to present (35, 68, 69). The dynamic, thermal, and magnetic energy (pressure) transported by the plasma directly affect the magnetic field environment of a planet, as a planetary magnetic field ($B_{planet}$, as measured at the surface equator) can be compressed inside the surface if it is not sufficiently strong (70). We use $B_{planet}$ throughout the text as a more intuitive stand-in for dipole moment (e.g., a Mercury planetary field with $B_{planet}$ = 200 nT has dipole moment of ~$3 \times 10^{19}$ $Am^2$).

Comparing the dynamic pressure, $P_{sw}$, and planetary magnetic field pressure gives insights into two fundamental TRM parameters: the origin of the field that magnetized the crust (whether the planetary dipole field or IMF is dominant at the surface) and the field's temporal variations (the periodicity of changes in the magnitude and direction of the magnetizing fields). These two aspects are linked because the timescale over which the dynamo and IMF remain steady differ greatly [e.g., ~$10^5$ years (39) and ~minutes (71) today for Earth, respectively]. If the dynamo field could only moderately stand off the solar wind, Mercury's surface would be embedded in the day-night asymmetric magnetospheric dipole field, experiencing different field strengths over its solar day. If the dynamo field was compressed within the surface, then the dayside surface would be embedded in the IMF which changed on timescales (~minutes) much shorter than the cooling rates for large crustal layers (~10 m has conductive cooling timescale ~1 y).

**Approach**. In this study, we quantify the magnetic field environment, thermal cooling, and magnetization processes at Mercury ~3.9–3.7 Ga ago to determine how the strength, origin, and temporal variability of surface magnetic fields affected the ability of crustal materials to acquire a TRM detectable by orbiting spacecraft. We consider the surface magnetic field environment at 40° N to compare with the strong MESSENGER crustal field signals and at 10° N as a prediction for future spacecraft measurements at lower latitudes not observed by MESSENGER [anticipating BepiColombo (28, 72)]. We consider TRM as the dominant form of magnetization due to its association with widespread surface volcanism (see above) and the higher field recording efficiency of TRM compared to SRM (40). We aim to answer the following questions. (1) Can the ancient IMF alone have been the magnetizing field responsible for the measured crustal anomalies? (2) If not, how strong must the ancient Hermean dynamo-generated surface magnetic field have been to magnetize the crust and what are the implications for the dynamo mechanism?

To model the magnetic field environment at Mercury, we first estimated the ancient solar wind and IMF conditions at the planet's orbit. For a detailed discussion on approximating the solar wind conditions at Mercury ~3.9-3.7 Ga ago, see section 2.2 of ref. (35). Based on the MLR, solar rotation, and solar magnetic field expected at this time (69), we found a range in ancient solar wind mass density of 200 amu $cm^{-3} \leq \rho \leq 5000$ amu $cm^{-3}$, velocity of $u \sim 1000$ km $s^{-1}$, temperature of $T \sim 2\times10^6$ K, and IMF of 300 nT $\leq B_{IMF} \leq 600$ nT, yielding a dynamic pressure range of 330 nPa $\leq P_{sw} \leq$ 8300 nPa (Fig. 1; see Table S2 for MHD simulation parameters). We next quantified how the solar wind parameters might change over the timescale that crustal rocks could acquire TRM. For conditions when the solar wind compresses the planetary dipole field within the surface, we considered both a uniform, steady field to place an upper bound on the magnetization acquired and a time-variable field to obtain a more realistic IMF-sourced magnetization. For the latter, we used Advanced Composition Explorer (ACE) spacecraft measurements (73) as a first-order approximate magnetizing field to demonstrate the effects of time-variability on TRM (hereafter referred to as the "scaled ACE" dataset; see Materials and Methods and Fig. S2). To move beyond the first-order approximations, we employed both numerical (Figs. S3, S4, S5, and S6) and analytical (Fig. 1) solutions to solve for the full magnetic field structure at the surface. The numerical solutions enabled the generation of a time series of the three-dimensional (3D) surface magnetic field ($B_{surface}$) at a given latitude that can be coupled with the crustal field calculations. To do these calculations, we performed 3D-magnetohydrodynamic (3D-MHD) simulations [see Supplementary Information] using the derived ancient solar wind conditions and a fixed $B_{planet}$ (see Materials and Methods for more information).

We then coupled these magnetic field histories with thermal simulations of cooling structures representing volcanic flows, intrusions, and impact-triggered heating events (Materials and Methods). Because cooling time decreases as structure size decreases, smaller structures are more likely to be able to record a relatively uniform magnetization even when magnetized by a time-varying field. However, once these structures become too small, there is not enough volume to produce a large crustal anomaly (i.e., 10 nT at 30-km altitude) even for strong, uniform source fields. As such, our fiducial heated geometry was a 50-km diameter cylinder of thickness 20-km, representing a large-scale structure near the maximum surface spatial scale for the ~30-km altitude MESSENGER measurements (note that the crustal field surface horizontal spatial scale resolvable from orbit is approximately that of the spacecraft altitude). This scale of magnetized material was chosen as a near upper limit on the possible generated crustal magnetic field from a reasonably sized volume, given that the strength of a field generated by a uniformly magnetized cylinder scales with volume (e.g., increasing the radius and thickness each by a factor of ~3× maximizes the crustal field at ~30-km altitude). A second reason for this choice of geometry is that its volume is similar to the likely size of the buried (10-40 km depth) magmatic sources inferred to generate much of the northern hemisphere anomalies (34). This thermal cooling model is hereafter referred to as "DSK50" (Fig. 2). We performed additional thermal cooling models for two smaller structures to test the trade-off between volume, cooling, and magnetic field timescale. These smaller structures were a cylinder of diameter 2-km and height 8-km representing a lava tube or vertical intrusion ("LT2") and a thin disk of diameter 20-km and thickness 1-km representing an effusive lava spreading event ("DSK20)."

Having determined the thermal and magnetic histories of these structures, we finally calculated the generated crustal fields which we then compared with MESSENGER measurements (see Results). We also compared our nominal MHD-coupled magnetization results with additional calculations incorporating the dynamo changing (reversing) in direction over the ~Ma cooling timescales (39). Although there is no conclusive evidence of Hermean dynamo reversals (32), it is important to discuss the effects of reversals on resultant crustal magnetization given their prevalence in dynamo models and observations (74). We utilize equation 2 in (39) to incorporate the diminishing effect of reversal rates ($f$ = 0.1-10 $Ma^{-1}$) on crustal anomaly magnitude based on the conductive cooling timescale of the heated geometries. From the findings of ref. (39), it is expected that lower reversal rates (e.g., ~0.1 $Ma^{-1}$) will have a smaller diminishing effect on the crustal anomaly magnitude when compared to higher reversal rates (e.g., 10 $Ma^{-1}$, Fig. S7). Furthermore, for a fixed reversal rate, larger geometries will have relatively weaker crustal anomalies when compared to those generated from non-reversing source fields (Fig. S7). We discuss the dynamo reversal rate effects in the context of our crustal field calculations in the Results section.

## Results

**Magnetization by an Ancient IMF.** As an initial probe of the surface magnetic field environment, we use equations (1-3) (Materials and Methods) to determine whether the IMF or dynamo field was dominant at the ancient surface. Figure 1 shows the analytical magnetopause distance, $R_{MP}$, as a function of ancient solar wind dynamic pressure, $P_{sw}$ [equation (3)]. For the present-day $B_{planet}$ ~200 nT dipole equatorial surface field (dipole moment ~$3\times10^{19}$ Am$^2$), the estimated magnetopause distance, $R_{MP} \lesssim 1$ Mercury radius, $R_M$, for the majority of the $P_{sw}$ range (i.e., 800-8300 nPa). Therefore, if the ancient dynamo were as weak as the present-day field, the surface would be exposed to the IMF and magnetospheric perturbed fields for nearly the full expected range of solar wind pressures.

We compared the magnetization of the LT2, DSK20, and DSK50 models magnetized by a temporally constant IMF of 500 nT with that for the same models magnetized by the natural, time-variable IMF as represented by the scaled ACE dataset with average magnitude, $B_{IMF}$ = 500 nT (Fig. S2). We found that the smaller volumes DSK20 and LT2 had ~20× and ~40,000× weaker remanent fields, respectively, compared to the DSK50 model. Additionally, we found that the crustal field at 30-km above the surface of the DSK50 magnetized volume was >1,000× weaker for the time-variable IMF (0.004 nT for $\chi_{TRM}$ = 0.02) compared to that from a 500 nT uniform, steady magnetic field. Moreover, the crustal field from the time-variable IMF (0.004 nT) was much weaker than the MESSENGER measurements of ~10 nT at 30-km altitude. This confirms our expectation that a time-varying field during cooling will result in a non-uniform, weaker magnetization that cannot account for the MESSENGER measured crustal fields.

**Magnetization by a Present-Day Strength Magnetospheric Field**. To model crustal magnetization with a realistic surface magnetic field timeseries reflecting the ancient solar wind interaction with a dipole field of present-day strength, we coupled our derived magnetic field timeseries from the set of steady-state 3D-MHD simulations with $B_{planet}$ = 200 nT as described in the Materials and Methods (Table S2, Fig. S4, and Movie S1) with the thermal cooling evolution models for LT2, DSK20, and DSK50 (Figs. 2 and 3). The lower end of the solar wind density, 200 amu cm$^{-3}$ (derived dynamic pressure of 330 nPa) was chosen for the 3D-MHD simulations. This value and all higher values of the mass density range (200 amu cm$^{-3}$ $\leq \rho \leq$ 5,000 amu cm$^{-3}$) resulted in compression of the dayside magnetopause at or within the surface, such that the IMF variability affected the surface magnetic field timeseries and diminished the magnitude of the magnetization. We found that the $B_{planet}$ = 200 nT dipole was too weak to magnetize the LT2 and DSK20 geometries to the levels measured by MESSENGER. Therefore, we focus below on the DSK50 magnetization results.

A magnetic field timeseries was sampled from the collection of 3D-MHD simulations at 40° N latitude (Fig. 2), the latitude at which some of the strongest crustal fields were measured (4). We found that the crustal field at 30-km generated from this geometry and magnetic history is only ~0.3 nT, ~3× weaker than a crustal field produced by a constant ~370 nT field (the present-day dipole field strength at 40° N when $B_{planet}$ = 200 nT). This calculation assumed the largest recording efficiency of known potential Hermean magnetic carriers, aubrite-like composition with $\chi_{TRM}$ = 0.02 [SI], and this value was used throughout the presented results unless otherwise noted. This ~3× weaker crustal field is a result of the surface compression of the dayside magnetopause, allowing for the variable IMF to influence the surface magnetic field for around half of the cooling time. Given that the measured long wavelength crustal field signals (29, 34) indicate that some of the northern crustal anomalies are generated from buried sources (10-40 km), we conducted an alternate calculation in which we buried the (top of) magnetized material at ~10-km depth; this further diminished the anomaly by a factor of ~3×. Similarly, a magnetic field timeseries was sampled at 10° N latitude as a prediction for future measurements (e.g., BepiColombo) (Fig. 3). At this latitude, the DSK50 magnetization resulted in a crustal field of ~0.11 nT at 30-km altitude, also ~3× weaker than for a constant ~210 nT source (dipole field at that latitude). In summary, at both 10° N and 40° N latitudes, we found that a dynamo field with present day strength at 3.9–3.7 Ga is highly unlikely to be able to magnetize the crust and reproduce the MESSENGER-measured signals. Therefore, the northern crustal anomalies strongly indicate a stronger ancient dynamo.

**Magnetization by a ≥10× Stronger Magnetospheric Field.** We repeated the crustal field calculations for the DSK50, DSK20, and LT2 geometries with magnetic field timeseries at 40° N and 10° N latitude for a 10× stronger dynamo field, $B_{planet}$ = 2,000 nT (dipole moment ~$3\times10^{20}$ Am$^2$) (Figs. 2 and 3). For the 40° N latitude magnetic field timeseries, the DSK50 geometry generated a ~3 nT field at 30-km, while a uniform, steady field produced a remanent field ~1.83× stronger (Fig. 2). This factor is smaller than the equivalent one for the present-day $B_{planet}$ because the stronger dipole field is better able to stand off the extreme solar wind, reducing the diminishing effects of the IMF variability and induced core-field day-night asymmetry ($B_{c,ind,}$ see Materials and Methods). For the 10° N latitude magnetic field timeseries (Fig. 3), we estimated a ~1.4 nT remanent field from the DSK50 geometry at 30-km altitude.

While these ~1-3 nT crustal fields are still lower than the strongest ~10 nT signals measured by the MESSENGER magnetometer (Fig. 4), increasing the volume of DSK50 to incorporate the entire crustal thickness (diameter of 50-km to 150-km and thickness 20-km to 70-km) would result in a factor of ~8× in the peak field strength. Therefore, we conclude that the observed 10 nT field at 30-km altitude can be generated by a 150-km diameter and 70-km (DSK150) thick layer magnetized by a dipole field with $B_{planet}$ = 2,000 nT. Even so, including the maximum crustal thickness of ~70-km [average crustal thickness is ~35-km (54, 55)] constitutes an extreme upper limit on a volume that could be uniformly cooled and magnetized (34).

**Adding Reversals.** As the conductive cooling timescales for DSK50 and DSK150 are >10 Ma, dynamo field reversals would diminish the resultant crustal field when compared to a steady dynamo source. For reversal rates, $f$ = 0.1–10 $Ma^{-1}$, the final anomalies are weakened by factors of ~1.2–20× for DSK50 and 4–80× for DSK150 (Fig. S7). To reach the 10 nT signals at 30-km altitude for $f$ > 1 $Ma^{-1}$, the DSK50 and DSK150 models would have to be magnetized by a $B_{planet}$ ~ 30,000 nT field (dipole moment ~$5\times10^{21}$ $Am^2$), ~100× stronger than the surface field at present and similar to Earth's present-day equatorial surface field. Assuming reversals are infrequent (i.e., baseline $f$ < 1 $Ma^{-1}$), the DSK150 upper limit on the magnetized volume implies a lower limit on $B_{planet}$ being ~2,000 nT. A field with $B_{planet}$ ~10,000 nT (~5× stronger) would allow for the DSK50 unit to reach ~10 nT at 30-km altitude, calculated by approximately linear scaling. This scaling is appropriate as such a field would have a subsolar magnetopause distance of ~3 $R_M$ and would thus be dominated by the steady dipole field at the surface.

## Discussion

**A New Minimum Paleofield Strength.** By combining recent advances in understanding Mercury's magnetic mineralogy with simulations of the ancient Hermean magnetic field environment, we have shown that both the ancient IMF and a source-field like that of the present-day Hermean dipole field are too weak to account for the strong northern-hemisphere crustal TRM. Even an enhanced surface dipole field of $B_{planet}$ ~2,000 nT (10× present-day) can only explain the strongest measured crustal fields with a magnetized volume of ~150-km diameter and ~70-km thickness, which approaches the maximum inferred crustal depth (54, 55). In fact, for source dimensions consistent with magnetization source-depth analyses (34) and the average crustal thickness (e.g., DSK50), aubrite-like magnetic compositions with $\chi_{TRM}$~0.02 require $B_{planet}$ ~10,000 nT to reach MESSENGER-level signals. This inference further requires that the dynamo field was steady over the ~80 Ma cooling time; if the field reversed with a frequency of >1 $Ma^{-1}$, similar to that for other terrestrial dynamos (39), the required surface field magnitude would need to be $B_{planet}$ >30,000 nT (Fig. S7). If it were found that some of the other potential magnetic carriers, like schreibersite, suessite, and rare occurrences of pyrrhotite in a shallow layer are present in the Hermean crust and have $\chi_{TRM} \gtrsim 0.1$ [SI], then a $B_{planet}$ ~2,000 nT field could magnetize the DSK50 source to the observed levels. Although previous work could not differentiate between $B_{planet}$ = 200 nT and $B_{planet}$ = 50,000 nT (4), these results indicate a new lower limit on $B_{planet}$ of >2,000 nT at ~3.9-3.7 Ga ago. Future studies of paleopole and unit age determination can quantify the Hermean dynamo reversal rate (32), further constraining the time-evolution of the ancient dynamo intensity.

**Mercury's Dynamo, Interior, and Thermal Evolution.** As previously introduced, the present-day Hermean dynamo-generated magnetic field has three central oddities: (1) it is weaker than predicted by dynamo scaling laws, (2) it has the geometry of an offset dipole (i.e., non-negligible multipolar power for a planet-centered spherical harmonic expansion), and (3) it is almost perfectly aligned with the axis of rotation. Because this study provides a determination of the ancient field intensity and not its geometry, we focus our discussion on the implications of our results for oddity (1) above. Our evaluation considers each dynamo mechanism as an endmember, noting that at present, Mercury's field may be generated from several of the processes previously described. The evolution of the dynamo magnetic field is intimately coupled to the thermal and compositional evolution of the planet, which have been investigated with modeling and observations to predict ancient surface volcanism, crustal thickness, heat flux, the core-mantle boundary radius, the inner-to-outer core size ratio, and the amount of planetary radial contraction over time [(75-81) and references therein]. Our finding that the Hermean dynamo field at the surface ~3.9-3.7 Ga ago was at least 10×, or possibly 50-100×, stronger than at present provides several insights into the interior evolution.

Firstly, we discuss this result in the context of interior properties. For dynamos governed by both the Elsasser number ~ 1 (8, 9) and available energy flux convection scalings (8), the dynamo region boundary heat flux, $Q_D$, would have been greater than the adiabatic heat flux, $Q_a$ [i.e., heat flux required to sustain convection, (75, 77)]. Such dynamos would generate a field at the core mantle boundary that is ~10-300× stronger than that inferred for the present, compatible with our findings (7, 74). In principle, if $Q_D$ was at its maximum >3.9 Ga ago and decayed with time, the magnetized units older than ~3.9 Ga would be expected to have stronger magnetizations, assuming the field geometry, field time variability, sizes of magnetized structures, and mineralogy are constant. Although a dynamo driven in part by core crystallization is possible for the present-day field (15, 19, 22, 26), such long-lived (i.e., since ~3.9 Ga ago) core crystallization would imply a large inner core at present, which is not supported by the current understanding of Mercury's planetary radial contraction history (82-85), magnetic field (86), and geodetic data (80, 87). Compatible with these present-day observations, the high heat flux driving the strong early dynamo likely did not allow for substantial inner core growth (14, 75). If the measured crustal fields are indeed indicative of surface structures that formed at ~500-800 million years (Ma) after planet formation, this provides a potential lower limit on the timing for the lifetime of this stronger, convection driven dynamo. Future measurements that are able to place tighter constraints on the dating (38) of these magnetized units can provide further information on the interior evolution (72).

Secondly, most of the dynamo models rely on a stably stratified layer that electromagnetically shields some of the magnetic field [effectively diminishing the short-wavelength and highly time-variable components (15)]. As a strong convection-driven dynamo would be powered by a sufficiently high heat flux (e.g., $Q_D > Q_a$ for a thermal convection dynamo), our results could be explained if the stably stratified layer invoked in these dynamo models [e.g., thin-shell convection (12, 13), convection under a thick stable layer with uniform heat flux (15), double-diffusive convection (21, 22), and stable layer with variable heat flux (27)] was thinner than at present (for a ~10× stronger dynamo) or did not exist at all in this period (for a ~50-100× stronger dynamo). Models of the interior evolution suggest that this stably stratified layer could have formed ~500-800 Ma after planet formation when the available heat flux in the dynamo region was no longer able to sustain well-mixed convection (75, 77). If the initial heat flux was able to drive whole core convection, then prior to ~3.9 Ga ago, the dynamo field might have been initially closer to ~50-100× the present value and then transitioned closer to the ~10× state when the available heat diminished and the stable layer formed (Fig. 5, Table S1). Upon the formation of the stable layer, the dynamo would have transitioned from the strong convection (dominantly thermal-driven) regime to being powered by thermochemical or chemical convection via chemical buoyancy [e.g., by iron-snow (18-20)]. It has been proposed that the Caloris basin formation event, likely occurring within this ~3.9-3.7 Ga ago window (88), could have suppressed the earlier strong convection dynamo and/or caused the dynamo to transition into one of the weaker field mechanisms (89). However, the role of impacts in perturbing dynamo activity remains speculative (90, 91).

Lastly, we evaluate the plausibility that the aforementioned dynamo mechanisms could have functioned in this ancient period or if they constitute later evolutionary stages of a once strong convection-driven dynamo (Fig. 5, Table S1). The thermoelectric dynamo (10, 11) likely would not have been a viable mechanism in the first ~1 Ga after the planet's formation due to the inferred high heat flux ($Q_D > Q_a$) and its inability to generate a strong planetary dipole field [the predicted field scales with local CMB temperature gradients and produces a three-orders of magnitude weaker field than as does a conventionally scaled convection dynamo (10)]. The solar-wind feedback (25, 92) dynamo also likely would not have been the dominant mechanism in this ancient period as this mechanism relies on an externally-induced field present at the dynamo-generating region that is similar in strength and/or stronger than the internally generated field. To demonstrate this limitation, the ~2,000 nT and ~10,000 nT surface dipole fields encountering the extreme ancient solar wind ($P_{sw}$ ~ $10^3$ nPa) would have subsolar magnetopause standoff distances of ~1.5 $R_M$ and ~3 $R_M$ and magnetopause current-induced fields at the core of only ~485 nT (~10% of field at dynamo region surface) and ~400 nT (~1% of field at dynamo region surface), respectively (Fig. 1; see Materials and Methods). Because the solar-wind feedback process requires the induced field at the dynamo region (i.e., core boundary) to be substantial [i.e., $\gtrsim$50% the field at dynamo region (25)] relative to the dynamo-generated field at the same location (24, 25), it likely could only have been viable when the dynamo-generated field was $\lesssim$10× the present-day strength.

Some of the thin-shell convection dynamo models (12, 13) posit that the dynamo action is confined to a thin convecting shell overlying a large, solid inner core. This model could be an evolutionary later stage of an early thick-shell convecting dynamo layer (which could predict a strong early field compatible with our findings) overlying a solid inner core smaller than that of today. However, it is likely that over ~3 Ga of core solidification required by such a model would result in more pronounced radial contraction than that observed [Figures 19.5 and 19.6 of (76)]. The thick-shell convection (14), iron-snow (18-20), and convection with variable heat flux (26) models represent possible present-day evolutionary states of an ancient, strong convection dynamo, yet suffer from not being able to simultaneously explain all three key features (e.g., the magnitude, axisymmetry, and geometry) of the present-day dipole field. Further, recent interior modeling suggests Mercury's core lacks the sulfur content necessary to form iron-snow layers (93). Conversely, the convection with a stable layer and laterally variable heat flux (27), double diffusive convection (22), and convection under a thick stable layer with uniform heat flux (17) dynamos are the only models at present able to capture all three prominent features of the present-day dynamo field. These dynamo mechanisms require small, solid inner cores (<0.2 $R_M$) and invoke stable layers of thickness 0.35-0.45 $R_M$ to mask the time-varying and higher order magnetic moments, both in line with the predictions of global radial contraction and interior evolution modeling (76). These models comprise thermal and/or chemical convection with a stably stratified layer that would have formed after the heat flux into the core lessened and core solidification commenced. As such, these mechanisms represent highly compatible evolutionary endmembers for a once stronger ancient convection dynamo.

**Outlook for BepiColombo.** The upcoming BepiColombo Mercury Planetary Orbiter (MPO) plans for an orbital altitude of ~200-km about 3 years after the start of its primary mission phase [likely early 2029 (72, 94)]. At these orbits, the crustal magnetization responsible for the ~10 nT field at 30-km altitude measurements will produce fields of only ~0.02 nT, which is below the estimated uncertainty of the expected magnetometry data (72). However, if MPO [or its sister spacecraft, the Mercury Magnetospheric Orbiter (95)] performs a final science campaign of low-altitude flybys (as did MESSENGER), its magnetometer will have the opportunity to survey both the measured (>35° N) and unmeasured (at low altitudes, <35° N) surface. These new magnetic measurements could be combined with the spectrometers and particle instruments to co-locate strongly magnetized regions with surface mineralogy characterization. In particular, measurements that identify the mineralogy and abundance of the dominant magnetic carrier by finding localized hotspots of iron-rich material (30, 31, 96, 97) could help break the degeneracies associated with translating crustal field measurements to paleofield estimates. Specifically, if one such large-scale magnetized region is known to be magnetically dominated by Fe or FeNi (44), then this can provide valuable insight into relating the measured crustal field to the ancient magnetizing field source. Additionally, if BepiColombo does measure a ~10 nT at 30-km altitude at lower latitudes (i.e., 10° N) co-located with ancient volcanic structures, this would strengthen the case for an ancient equatorial surface dipole field of >2,000 nT. Combining future BepiColombo studies of crustal fields, surface mineralogy, and surface age, with rock magnetic analyses of Mercury analog samples can help break the degeneracy of magnetic modeling of the surface and provide more precise constraints on the magnitude and temporal evolution of the dynamo.

## Materials and Methods

**Ancient Solar Wind and IMF at Mercury**. The flowing solar wind can be described by the MHD approximation over Hermean radius length scale using the average magnetic fluid quantities: mass density ($\rho$), bulk velocity vector ($\boldsymbol{u}$), average temperature ($T$), and IMF vector ($B_{IMF}$). These fluid parameters are governed by solar properties like the MLR, rotation rate, and surface magnetic field. Based on the ranges of MLR, solar rotation, and solar magnetic field expected in this ancient period, we arrive at the solar wind and IMF conditions at Mercury as shown in Table S2 and derived in section 2.2 of (35). For all runs, the solar wind mass density (ρ), temperature ($T$), velocity in $X$-direction ($u_X$, Fig. S1) were fixed at $\rho$ = 200 amu $cm^{-3}$, $T = 2\times10^6$ K, and $u_X$ = 1000 km $s^{-1}$, respectively. The mass density value utilized represents the lower limit on the possible range of this parameter for the Sun ~3.9-3.7 Ga (35). For this mass density and higher values ($\rho$ ~ 5,000 amu $cm^{-3}$), the $B_{planet}$ = 200 nT surface dipole field cannot sufficiently stand-off the solar wind on the dayside, effectively allowing for the IMF variability to influence the resultant surface field (Fig. 1). The temperature does not have a direct effect on the resultant surface magnetic field. The velocity and IMF likely did not vary by more than a factor of ~2× (35), thus the resultant surface field calculation would not vary drastically over the ranges of these parameters.

Importantly, the IMF varies both in magnitude and direction over a wide range of timescales (98) and the MESSENGER magnetometer found that the upstream IMF direction changed on the order of ~tens of minutes (71). Because MESSENGER only spent four years in orbit and had intermittent solar wind coverage (4), we instead used the total 22 year, hourly magnetometer data from the Lagrange point 1 solar wind monitor ACE spacecraft (73), and scaled the data to the ~500 nT ancient IMF at Mercury ("scaled ACE" dataset). While this scaled ACE dataset does not capture higher occurrence-rate, energetic, transient field enhancements generated by the young sun [e.g., near-daily coronal mass ejections (CMEs) (99, 100)], such variability occurs on <1 hour timescales, far shorter than the million-year cooling timescale of the ~tens of kilometer-scale structures that acquire TRM. Adding in additional short-timescale variability would therefore tend to average out over the cooling process, and if anything, further reduce the net magnetization. As further justification for using the scaled ACE dataset for the ancient Mercury upstream conditions, we also note that the average Parker spiral angle at Earth is ~45°, meaning that in the ecliptic plane, the relative average magnitude of the sunward (radial) and azimuthal (angular) IMF components are roughly equal (101). For the range of spin rates for the young Sun, $2\Omega_\odot$ and $5\Omega_\odot$ (where $\Omega_\odot$ is the current spin rate) (69), the Weber-Davis solar wind model in (102) predicts a spiral angle of ~40°, indicating that the scaled ACE dataset provides a good approximation for the relative magnitude of the sunward and azimuthal ancient IMF components at Mercury (Fig. S2). Given the difficulty of solving for solar wind and IMF propagation in three dimensions and given that the out-of-ecliptic IMF component was the largest magnitude component for only 20% of ACE and MESSENGER measurements (71), we took the out-of-the-ecliptic plane component as measured by ACE for the time series.

**Mercury Surface Magnetic Field Environment**. With the derived solar wind and IMF conditions at Mercury's orbit, we employed both analytical and numerical solutions to solve for the 3D magnetic field structure at the surface needed to model the magnetization and generated crustal fields. As a starting point, we assumed the ancient dynamo-generated planetary field has the present-day dominant dipolar geometry [with the dipole center shifted ~500-km northward along the spin axis (3)]. To first order, the ability for a planetary magnetic field to stand off the solar wind at its dayside magnetopause can be estimated from the balance of the pressure from the planetary magnetic field and the dynamic pressure of the streaming solar wind (70).

$$\rho u^2 = f^2 {B_{planet}}^2/(2\mu_0) \qquad (1)$$

$B_{planet}$ is the planetary field at the surface magnetic equator, $f$ represents the amplification factor of the magnetic field at the magnetopause due to the Chapman-Ferraro currents [usually taken to be $f$~1.44 (103)] and $\mu_0$ is the magnetic vacuum permeability.

However, for the ancient Mercury environment, there are additional factors that can affect the magnetopause location and change the surface magnetic field environment. Firstly, the pressure balance should include the IMF pressure as $B_{IMF}$ was likely of the same order as the present-day Mercury equatorial dipole field strength. Secondly, due to the large metallic core (~80% of the Hermean radius), secondary fields can be induced in the core ($B_{c,ind}$, as measured at the surface magnetic equator) from the compression of the planetary magnetic field, increasing the overall magnetic pressure exerted by the magnetosphere (23, 104, 105). Lastly, although difficult to incorporate in a simple, analytical manner, the magnetopause location can be further moved towards the planetary surface due to magnetic reconnection when the dayside IMF is anti-parallel to the planetary field, commonly referred to as "reconnection-driven erosion" (105). However, previous simulations under CME conditions suggest that "reconnection-driven erosion" reduces the standoff distance by only ~10% relative to nominal pressure-balance expectations [equation (1)] (105), which is small compared to effects from the possible range of ancient solar wind dynamic pressures (which can range from ~100 nPa to ~9000 nPa). We neglect the IMF pressure as it only is relevant for <20% of the scaled ACE dataset, when $B_{IMF}$ is parallel to $B_{planet}$ at the magnetopause (also noting that the thermal pressure is typically <10% of the dynamic pressure). Incorporating the core induction ($B_{c,ind}$) into the pressure balance formulation, we see that

$$\rho u^2 = f^2 (B_{planet} + B_{c,ind})^2/(2\mu_0) \qquad (2)$$

where $B_{c,ind}$ can be estimated using the approximations in equation 23 of (92). Finally, we arrive at the formulation for the subsolar magnetopause standoff distance, $R_{MP}$ (measured from the surface in units of Mercury radius, $R_M$), the boundary between the regions dominated by the solar wind and magnetosphere:

$$R_{MP} = \left[\frac{f^2 \left(B_{planet} + B_{c,ind}\right)^2}{2\mu_0 \rho u^2}\right]^{1/6} \qquad (3)$$

The value of $R_{MP}$ provides insight into the dominant magnetic field at the surface. If $R_{MP} \lesssim 1.0\ R_M$, then the magnetic field environment at the dayside surface will experience the ambient IMF and perturbed magnetospheric field (e.g., from magnetic field amplification at the core, dayside-nightside asymmetries, and magnetospheric cusps), which varies on timescales governed by the IMF variability (~minutes) and Mercury rotation period (~176 days). Conversely, if $R_{MP}$ > 1.0 (and especially if $R_{MP}$ >> 1.0), the surface magnetic field will be dominated by the planetary dipole field, which varies over dynamo variability timescales [~millions of years (39)].

**Surface Magnetic Field Timeseries Derivation (Fig. 6)**. This magnetopause calculation [equation (3)] gives an estimation of the dominant field source at the surface on the dayside. To obtain a realistic time series of the 3D surface magnetic field ($B_{surface}$) that can be coupled with the crustal field calculations, we performed 3D-MHD runs with varying IMF directions, magnitudes, and solar wind mass densities for a fixed $B_{planet}$ (Table S2). As it is computationally infeasible to run a 3D-MHD model for ~$10^6$ years of simulated time with ~1 minute temporal resolution, we performed 10 runs to steady-state with varying IMF directions, magnitudes, and solar wind mass densities and for a fixed $B_{planet}$ (see Table S2). We defined two coordinate systems for the derivation of the magnetic field timeseries (Fig. S3). The first is a planetocentric, Mercury-Sun-orbital Cartesian coordinate system where +*X* points from the Sun to Mercury, +*Y* points in direction of planetary orbital motion, and +*Z* completes the right-handed system. The second is a Sun-fixed (i.e., does not rotate with the planet), planetocentric latitude-longitude ($\theta$, $\phi$) coordinate system in which latitude is measured from the equator (0°) and longitude is measured from the fixed subsolar meridian (0°).

Using these coordinate systems, we mapped the surface magnetic field from the spatial domain to the time domain, taking the longitude coordinate ($\phi$) as time ($t$) for the magnetic field seen at a surface point at some latitude ($\theta$) over the Hermean day. Using longitude as a proxy for the time evolution of the magnetic field at a set latitude is possible due to the axisymmetry of the Hermean planetary field [(3), Figs. S3, S4, and S6]. For example, this means that for constant upstream solar wind and IMF conditions, the same subsolar surface point will experience the same magnetic field at time = 0 and time = 176 days. Incorporating the solar wind and IMF variability, we created this time series by sampling the normal distributions defined by the mean and standard deviation of the magnetic field components from the 10 simulations, also taking into account the relative likelihood for a set of upstream conditions. For this sampling, 80% of draws were with the IMF conditions matching the Parker spiral directionality and 20% of the draws represented conditions when the out-of-ecliptic plane component was dominant, reflecting scaled ACE IMF distributions. Drawing from the derived normal distributions allows for differing surface fields representing variation in the solar wind parameters and returns negligibly different (<5%) resultant crustal field calculations compared to directly sampling from the 10 simulation magnetic field solutions (Fig. S6). Lastly, we transformed the magnetic field from the planetocentric (*X, Y, Z*) to the local tangent plane coordinate system (*East*, *North*, and *Up*) for the thermal cooling and magnetization modeling (Fig. S3). This provides a method to describe the surface magnetic field over the Hermean day (i.e., as the planet revolves around its spin-axis), allowing for the calculation of the surface TRM generated crustal fields. A depiction of the magnetic timeseries derivation from 3D-MHD surface maps can be seen in Figures S4, 6, and movie S1.

**Thermal Cooling, Magnetization, and Crustal Field Modeling**. The thermal cooling model was performed using the 3D finite element method built with the *deal.II* library (106) as implemented in (107), assuming heat transport is purely conductive. The geometries of heated structures emulate surface volcanic flows (thin disks or rectangular sheets), intrusions (long cylinders) or craters (near-equilateral cylinders) with fixed temperature boundary conditions at their tops (planetary surface) and bottoms and zero heat flux boundary conditions at their sides. Temperatures for these structures were set to ~1700 K, consistent with the estimated average liquidus for Mercurian lavas (49). Heating was superimposed onto a three-layer background temperature profile assuming a surface temperature of 440 K and a heat flux of 25 $mW/m^2$, approximately reproducing the expected equilibrium temperature profile in Mercury's interior at ~3.0 Ga [(78, 108), Table S3]. Since we restricted our simulations to shallow regions of the crust and lithospheric mantle, we neglected the effects of mantle convection. The total simulation time was set to ~100× the estimated conductive cooling time of the heated geometry [see Methods in (39)]. The cooling simulation timestep was set to between 0.1 and 1 year, increasing by 5% at each timestep due to the shortest allowable timestep increasing as the temperature gradients smooth over time (39, 109).

The thermal models were then used to calculate magnetization acquisition at each timestep, using the kamacite (an example FeNi-metal phase found in aubrites, see Introduction) thermal unblocking spectrum [magnetization acquisition and loss as a function of temperature (110)] to estimate the fraction of magnetization blocked since the previous timestep. The fraction of magnetization at each timestep was set by the surface magnetic field vector at that time (taking the average vector over the time step if the cooling timestep exceeds the magnetic field timeseries cadence). At the completion of the thermal cooling and magnetization model, we then calculated the crustal field from the magnetized structure using the rectangular prism method in (111) to sum the contributions from each voxel (3D grid cube).

To build intuition for interpreting the calculated crustal fields, we discuss the relationship between the diameter (*D)* and thickness (*H*) of a cylinder and its remanent field, *B*, at a given spacecraft altitude (*a*). We take three limiting cases for the diameter: *D* << *a*, *D* ~ 2*a*, and *D* >> *a*. For *D* << *a*, the cylinder's crustal field will be low (approaching 0 nT as *D* → 0-km) as there is not enough volume of magnetized material, even for *H* ~ *a*. *D* ~ 2*a* produces the maximum generated crustal field; this will asymptote to a maximum value with increasing thickness (e.g., when $H \gtrsim a$) as this adds more magnetized material at greater distances from *a* which negligibly add to *B*. For *D* >> *a*, the cylinder's field will approach zero for all *H*, the so called zero-field infinite plane solution (112). These general relationships between volume parameters and generated crustal field can be seen in Figure S9 with fixed *B* = 10,000 nT (along the cylinder long-axis or *H* direction) and $\chi_{TRM}$ = 0.02 [calculated with *magpylib* (113)].

## Acknowledgments

The authors thank Hao Cao for his keen insight and discussions on the implications of the dynamo and interior evolution. ISN and BPW thank the NASA FINESST (grant 80NSSC23K1363) program. Resources supporting this work were provided by the NASA High-End Computing (HEC) Program through the NASA Advanced Supercomputing (NAS) Division at Ames Research Center. The authors acknowledge the MIT SuperCloud and Lincoln Laboratory Supercomputing Center for providing (HPC, database, consultation) resources that have contributed to the research results reported within this paper/report.

**Funding:**

NASA FINESST Grant #80NSSC23K1363 (ISN, BPW)

**Data and materials availability:** All data needed to evaluate the conclusions in the paper are present in the paper, the Supplementary Materials, and/or can be found at *TBD-Zenodo Link*. The BATS-R-US model is open source as part of the Space Weather Modeling Framework (http://github.com/SWMFsoftware). The BATS-R-US version used for the MHD simulations in this study is available at https://doi.org/10.5281/zenodo.14545143. All custom code used in this work to calculate volume cooling histories and perform magnetic calculations is available at https://github.com/ssteele1111/BasinCoolMag.

**List of Supplementary Information Includes:**

Table S1-S3

Figures S1-S9

Movies S1

**Figures and Tables**

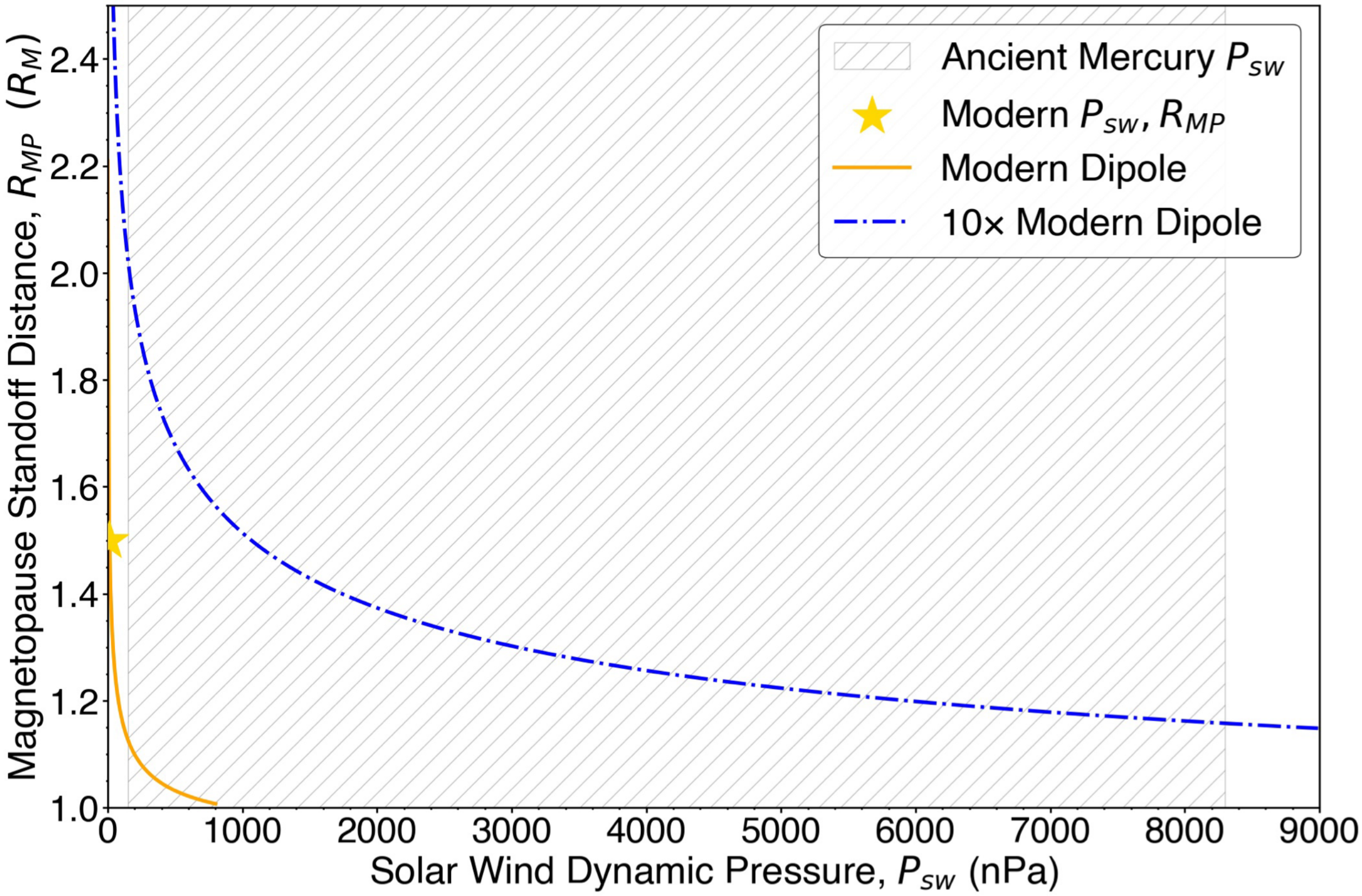

**Figure 1. Analytic calculation of the subsolar magnetopause standoff distance for various Hermean fields.** Shown is the Mercury magnetopause location ($R_{MP}$) as a function of solar wind dynamic pressure ($P_{sw}$) ~3.9-3.7 Ga ago. The ancient Mercury dynamo field would have to have been ~10× stronger than today to adequately standoff the solar wind (for most of the total range of $P_{sw}$) and be dominated by the dipole field at the dayside surface. The surface magnetic field is influenced by IMF variability for conditions leading to $R_{MP}$ ~ 1, which diminishes the possible magnitude of magnetization. The orange curve shows the analytical magnetopause location for a $B_{planet}$ = 200 nT surface dipole field with the additional effects of the magnetopause currents (Chapman-Ferraro) and the field induced in the core by the magnetopause currents. The blue curve is the same model as orange, but for the 10× stronger dipole field of $B_{planet}$ = 2,000 nT. The diagonally shaded background shows the range of $P_{sw}$ at the orbit of Mercury ~3.9-3.7 Ga ago. The yellow star represents the nominal present-day conditions, $P_{sw}$ = 10 nPa and $R_{MP}$ = 1.5 $R_M$, which vary negligibly (i.e., less than the size of the yellow star) in magnitude compared to the ancient solar wind pressure range.

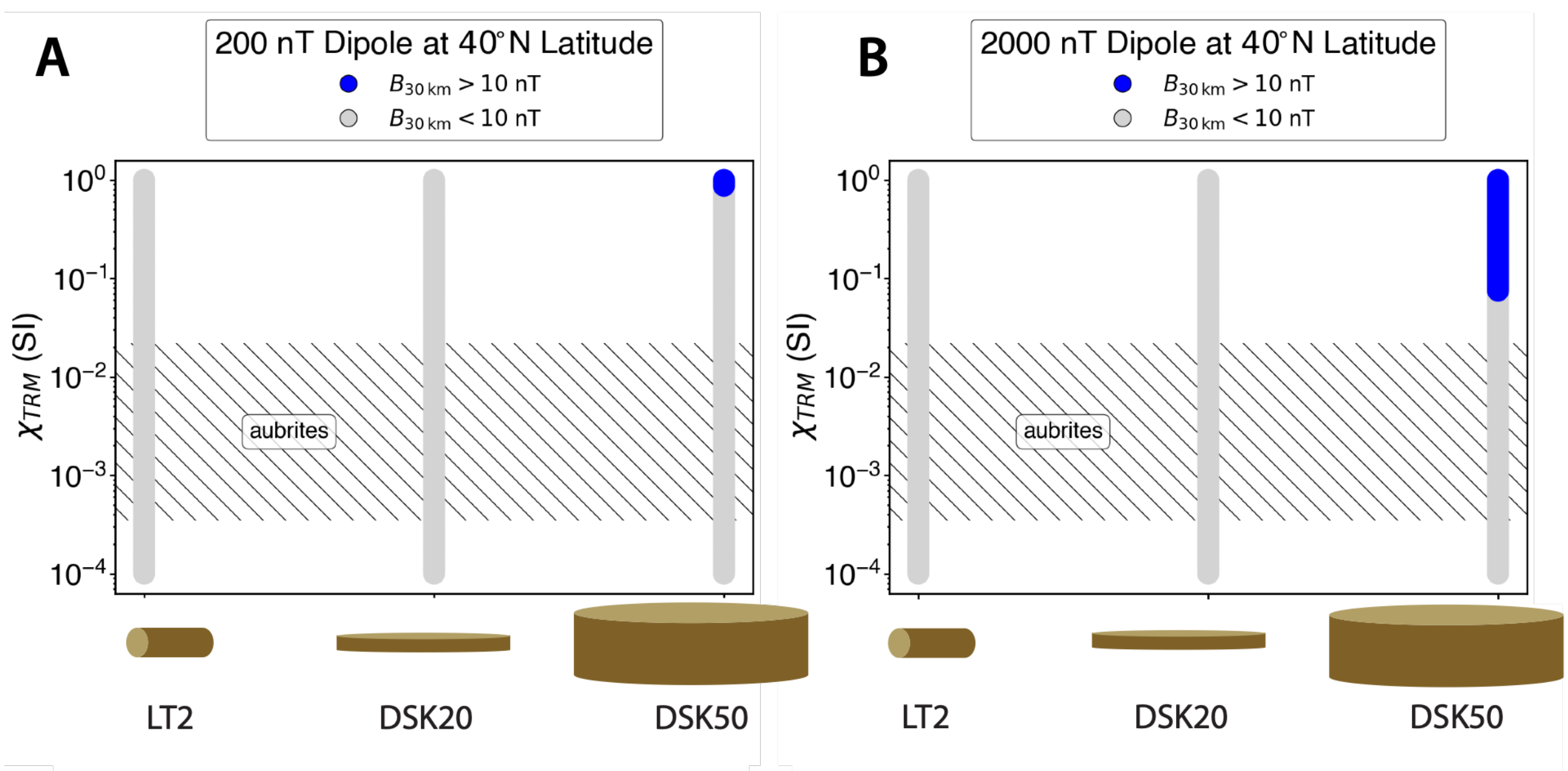

**Figure 2. Crustal fields from magnetized volumes in 200 nT and 2,000 nT equatorial surface dipole fields at 40° latitude.** Shown is the TRM recording efficiency ($\chi_{TRM}$, vertical axis) for different magnetized volumes (LT2, DSK20, and DSK50) that produce crustal fields (grey) <10 nT at 30-km altitude and (blue) >10 nT at 30-km altitude (i.e., MESSENGER-level signals). The magnetizing fields for these volumes are taken from the surface field timeseries at 40° latitude derived from the simulations of solar wind interaction with the (**A**) $B_{planet}$ = 200 nT and (**B**) $B_{planet}$ = 2,000 nT equatorial surface dipole fields. The reference $\chi_{TRM}$ values are taken to be that of the aubrites (hatched region). For these volumes and TRM recording efficiencies, the $B_{planet}$ = 200 nT dipole surface field timeseries is unable to produce crustal fields as measured by MESSENGER. Similarly, the $B_{planet}$ = 2,000 nT dipole field also struggles to account for these strong measured fields, but we note that a 150-km diameter and 70-km thick layer could produce the strong crustal fields (Fig. 4), though this is likely an unrealistically large layer. Shape sizes for LT2, DSK20, and DSK50 not to scale.

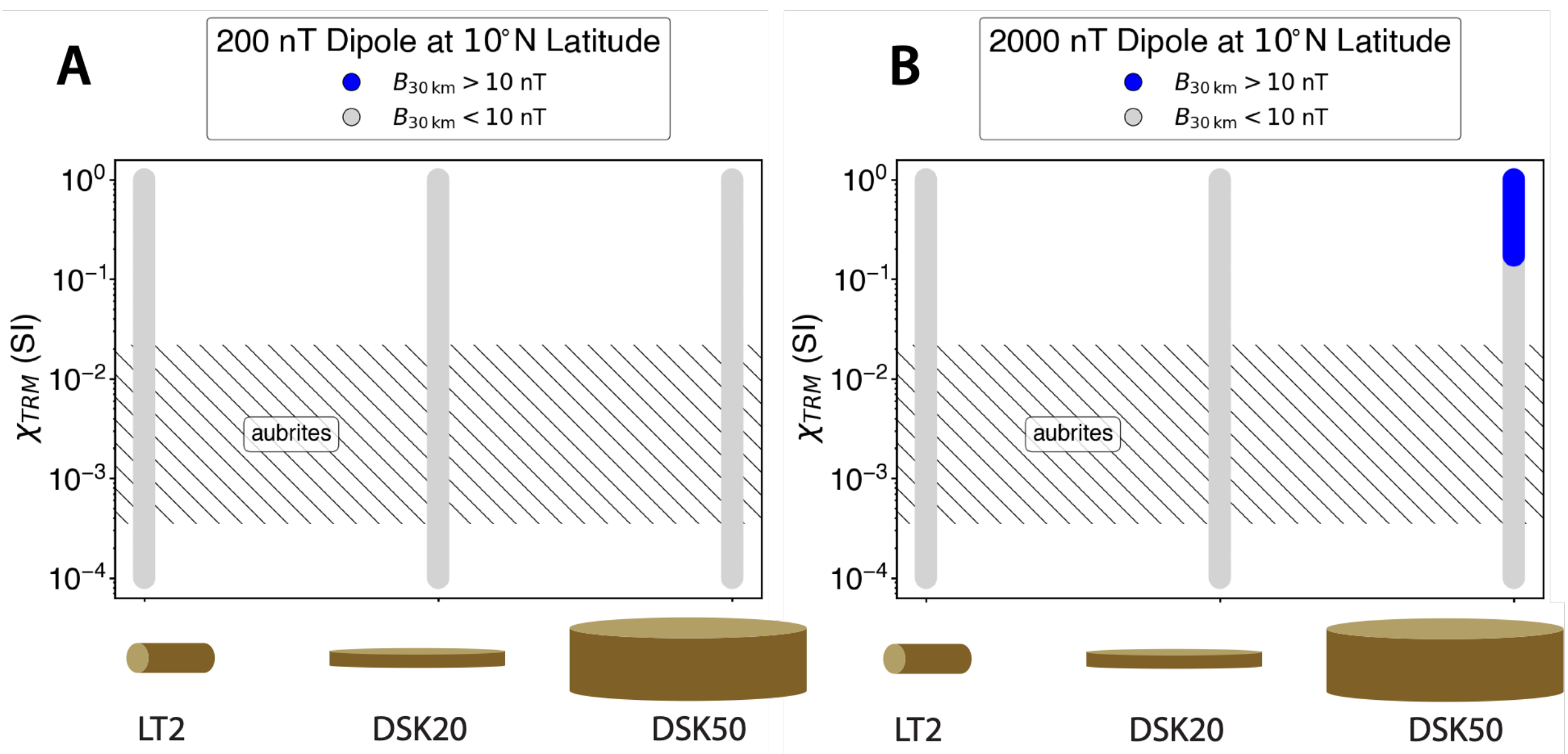

**Figure 3. Crustal fields from magnetized volumes in 200 nT and 2,000 nT equatorial surface dipole fields at 10° latitude.** Shown is the TRM recording efficiency ($\chi_{TRM}$, vertical axis) for different magnetized volumes (LT2, DSK20, and DSK50) that produce crustal fields (grey) <10 nT at 30-km altitude and (blue) >10 nT at 30-km altitude (i.e., MESSENGER-level signals). The magnetizing fields for these volumes are taken from the surface field timeseries at 10° latitude derived from the simulations of the solar wind interaction with the (**A**) $B_{planet}$ = 200 nT and (**B**) $B_{planet}$ = 2,000 nT surface dipole fields. The reference $\chi_{TRM}$ values are taken to be that of the aubrites (hatched region). For these volumes and TRM recording efficiencies, both the $B_{planet}$ = 200 nT and $B_{planet}$ = 2,000 nT dipole surface field timeseries are not able to produce crustal fields as measured by MESSENGER. The generated crustal fields at 10° latitude for both dipole field timeseries are ~2-3× weaker than those generated by the same conditions at 40° latitude. Shape sizes for LT2, DSK20, and DSK50 not to scale.

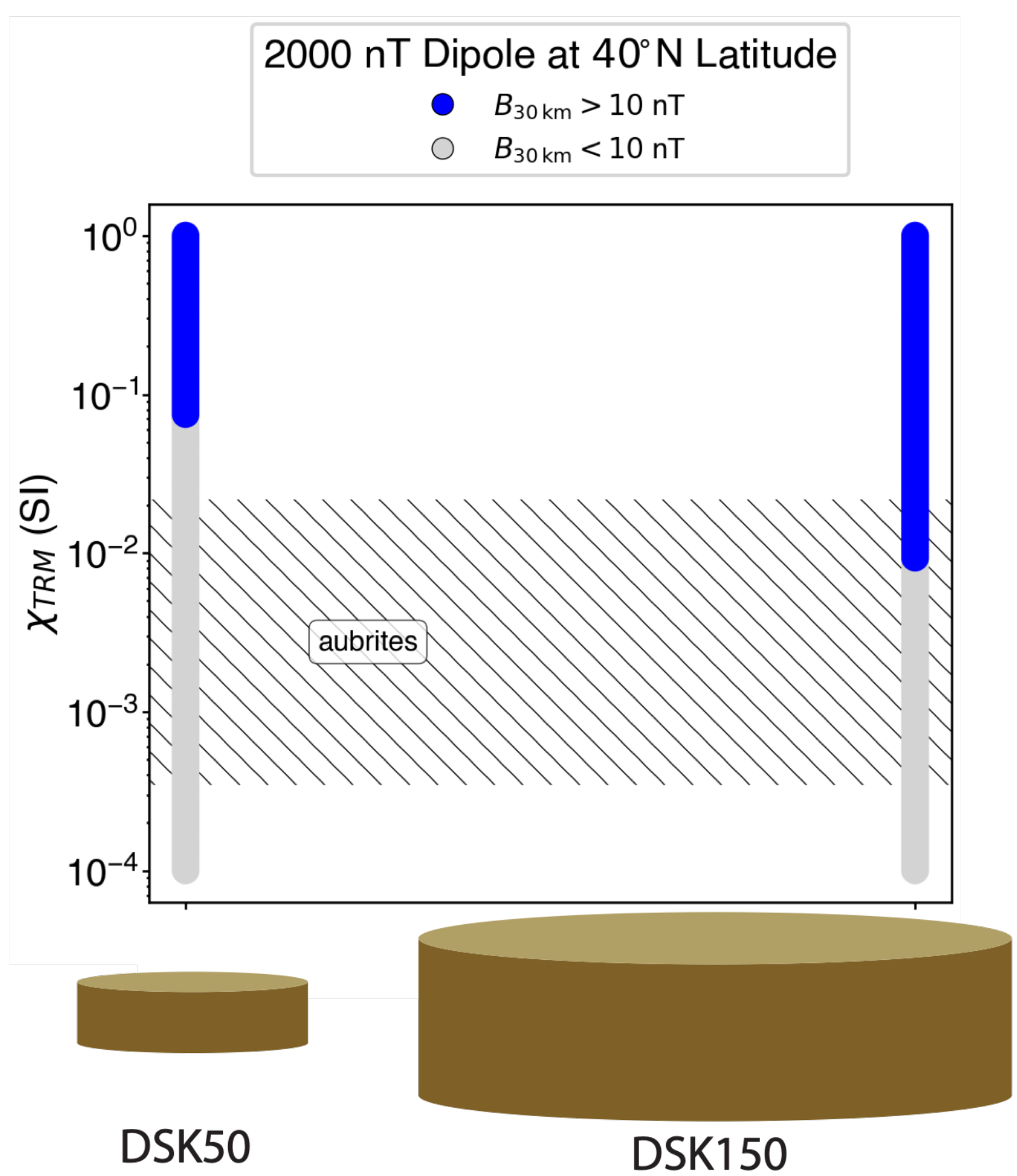

**Figure 4. Generated crustal fields from two magnetized volumes in 2,000 nT equatorial surface dipole field at 40° latitude.** Shown is the TRM recording efficiency ($\chi_{TRM}$, vertical axis) for different magnetized volumes (DSK50 and DSK150) that produce crustal fields (grey) <10 nT at 30-km altitude and (blue) >10 nT at 30-km altitude (i.e., MESSENGER-level signals). The magnetizing fields for these volumes are taken from the surface field timeseries at 40° latitude derived from the solar wind interaction with the $B_{planet}$ = 2,000 nT dipole field. The reference $\chi_{TRM}$ values are taken to be that of the aubrites (hatched region). While the 50-km diameter disk (DSK50) cannot account for the strong measured crustal fields, increasing the diameter and thickness by a factor of ~3 each allows for the 150-km diameter disk (DSK150) to generate the measured crustal fields. While this magnetization is possible, the DSK150 thickness is that of the maximum crustal thickness, thus representing an extreme upper limit for a magnetized volume. Scaling the crustal fields for the DSK50 by a factor of 5 to represent being sourced by a surface dipole field of $B_{planet}$ ~10,000 nT would match the measured crustal fields with aubrite composition magnetic recorders (see Discussion). Shape sizes for DSK50 and DSK150 are not to scale.

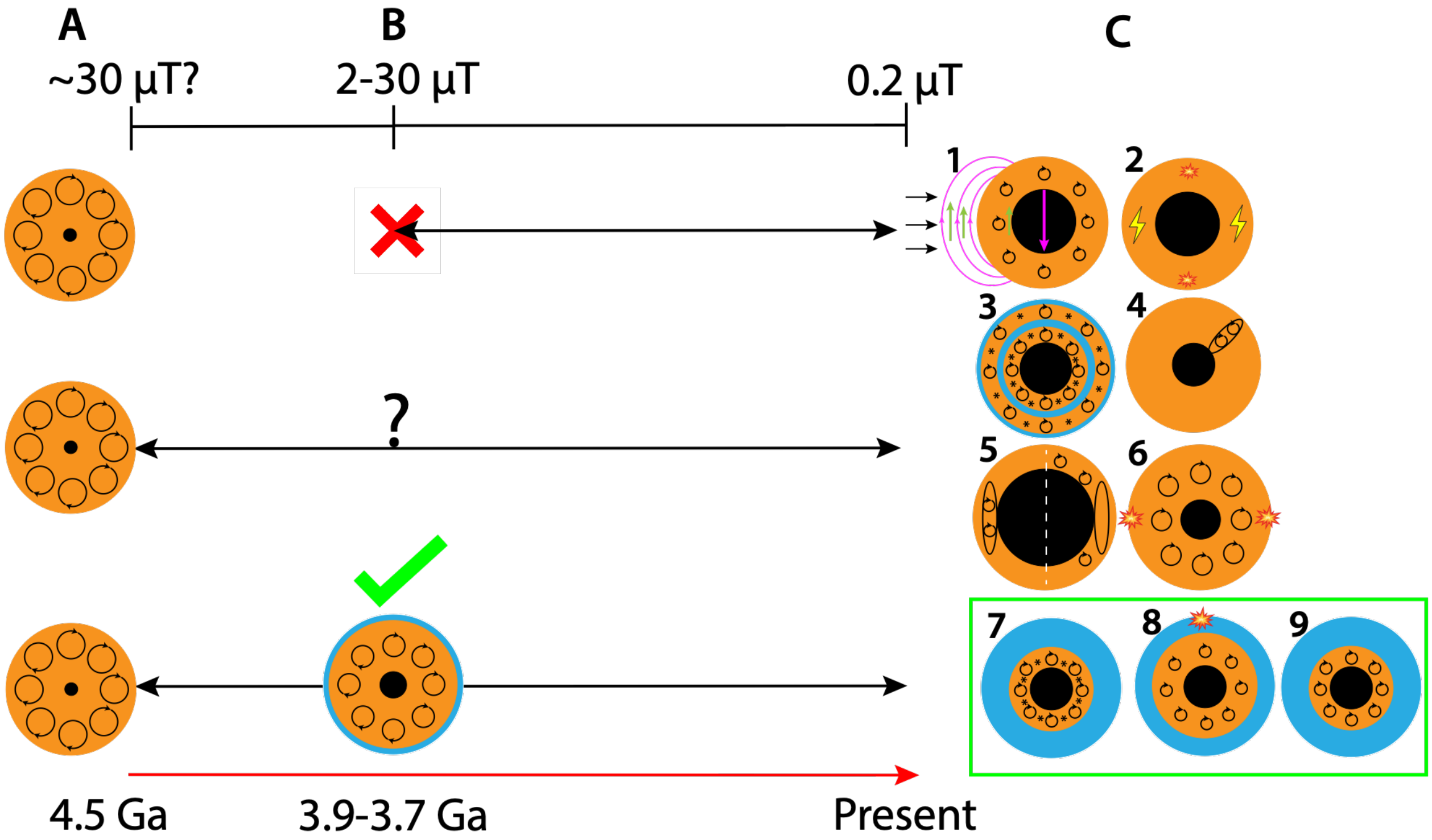

**Figure 5. Possible evolutionary paths of proposed Hermean dynamo mechanisms.** Shown are cartoon depictions of the mechanisms proposed to explain the ancient and present-day Hermean dynamo-generated magnetic field as a function of time. The red arrow shows the temporal axis, divided into three time-periods, (column **A**) planet formation 4.5 Ga ago, (column **B**) ancient crustal magnetization 3.9-3.7 Ga ago, and (column **C**) present period. The top horizontal axis shows the corresponding surface equatorial magnetic field magnitude for each time period, (column **A**) ~30 $\mu$T, (column **B**) 2-30 $\mu$T, and (column **C**) 0.2 $\mu$T, given by the findings of comparing our magnetic modeling to observations. Each dynamo cartoon depicts the total core with the black circle representing the solid inner core, the orange representing the liquid region, circular black arrows showing convection, and blue denoting layers stable to convection. The evolutionary paths start from the (column **A**) inferred strong convection dynamo which could have generated a 30 $\mu$T field based on Elsasser number and available energy flux scaling. The top evolutionary track shows the (column **C**) two dynamo mechanisms, (**1**) solar wind feedback (23-25) and (**2**) thermoelectric (10, 11), which are likely not able to explain (red "x") the stronger inferred planetary field, 2-30 $\mu$T, 3.9-3.7 Ga ago. The middle evolutionary track shows four dynamo mechanisms, (**3**) iron-snow (18-20), (**4**) thick-shell convection (14), (**5**) thin-shell convection (12, 13), and (**6**) convection with variable heat flux (26), which are possible products of a once stronger convection mechanism (question mark in column **B**) but either cannot explain all aspects of the present-day field or are not compatible with interior evolution modeling and observations. The bottom evolutionary track shows three dynamo mechanisms, (**7**) double diffusive convection (22), (**8**) convection with a stable layer and laterally variable heat flux (27), and (**9**) convection under a thick stable layer with uniform heat flux (17). These three mechanisms are highly compatible (green check mark) with the proposed evolutionary path; a strong convection dynamo, 4.5-3.7 Ga ago, generated a 2-30 $\mu$T magnetic field which then (third row cartoon in column **B**) weakened in strength due to the diminishing of the available heat flux into the core and the possible formation and growth of a shielding stable layer.

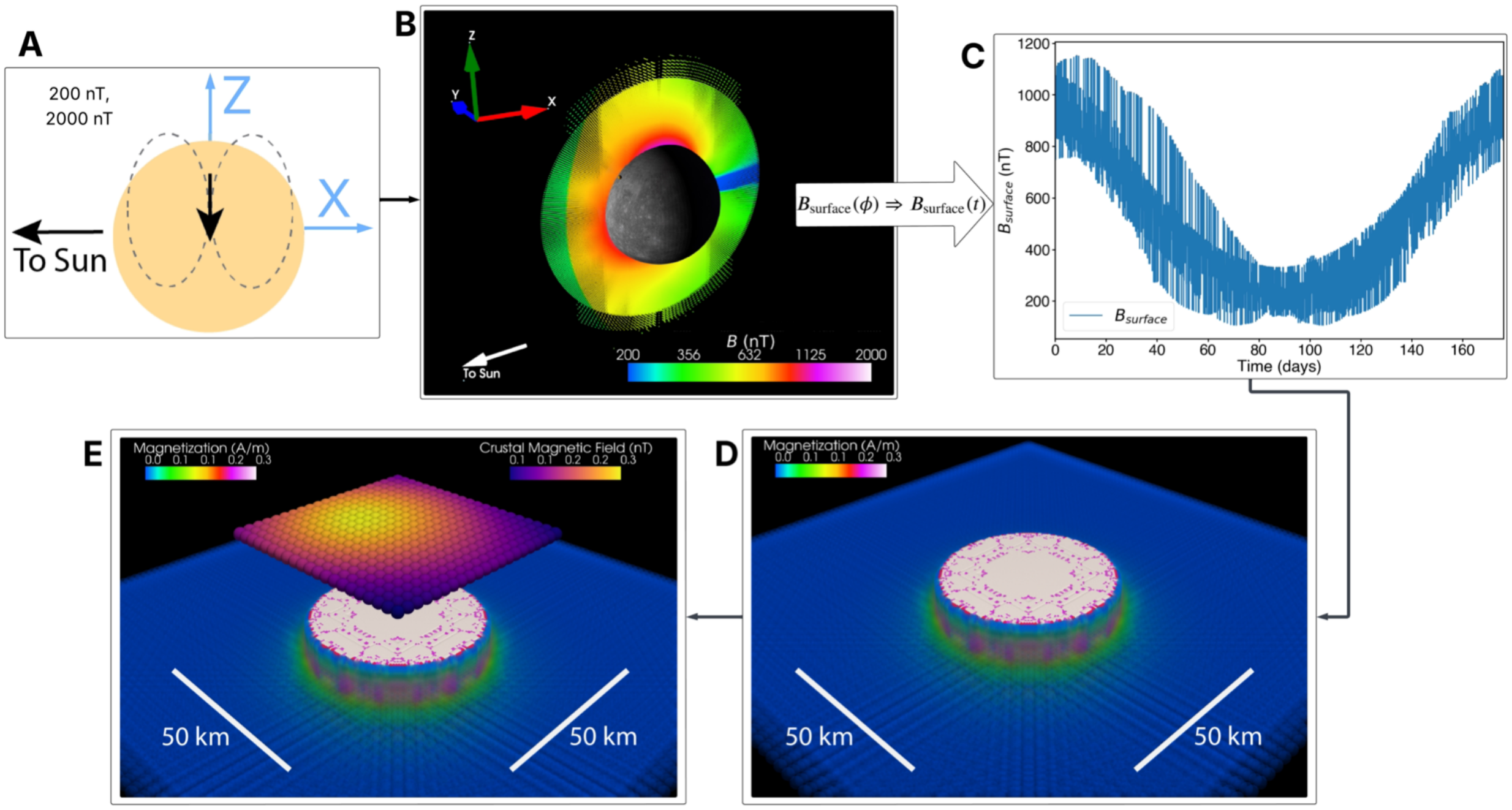

**Figure 6. Computational pipeline starting with the 3D-MHD simulation and ending with crustal magnetic field calculations.** Our workflow starts with the (**A**) input dipole field geometry and magnitude and solar wind configurations to produce the (**B**) 3D-MHD simulations of the surface magnetic field. This process is repeated for each of the 10 solar wind configurations for each separate dipole field ($B_{planet}$ = 200 nT and 2,000 nT) defined in Table S2 (movie S1). With the magnetic field solutions from the 10 solar wind configurations interaction with the dipole field, we then randomly sample from the MHD magnetic field solution at a given latitude over all longitudes to generate a magnetic field vector timeseries (**C**). This process essentially takes the surface magnetic field over the 360° of longitude and maps it to a timeseries, allowing us to use longitude as a proxy for time as the planet rotates (Fig. S4). This process allows us to use a surface-fixed magnetic field timeseries to self-consistently solve for the magnetization during the thermal cooling model calculations (**D**). These cooling geometries are the “DSK50” (cylinder of 50-km diameter and 20-km thickness representing a large crater or sheet), “DSK20” (20-km diameter and 1-km thin disk an effusive lava spreading event), and one “LT2” (cylinder of 2-km diameter and height 8-km representing a lava tube or vertical intrusion). Once we have calculated the magnetization for each of the volumes, we calculate the crustal fields (**E**), which we then compare with the MESSENGER measurements (Fig. S5).

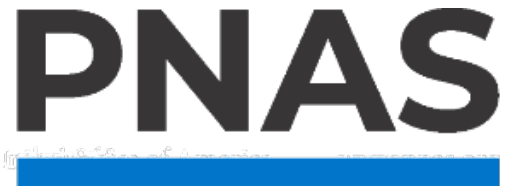

# Supporting Information for

# Mercury's Crustal Magnetization Indicates a Stronger Ancient Dynamo

Isaac S. Narrett[1]*, Benjamin P. Weiss[1], Sarah C. Steele[2], John B. Biersteker[1]

[1]Department of Earth, Atmospheric, and Planetary Sciences, Massachusetts Institute of Technology, Cambridge, MA, USA.

[2]Department of Earth and Planetary Sciences, Harvard University, Cambridge, MA 02138, USA.

*Corresponding author. Email: Isaac S. Narrett, narrett@mit.edu

**This PDF file includes:**

Supporting Information Text

Table S1-S3

Figures S1 to S9

Legends for Movies S1

**Other supporting materials for this manuscript include the following:**

Movies S1

**Supporting Information Text**

**Calculating Thermoremanent Magnetization (TRM) Recording Efficiency for Mercury Magnetic Mineralogies**

Here, we estimate the TRM recording efficiency of candidate Mercury crustal magnetic mineralogies from laboratory measurements of saturation remanence, $M_{RS}$ [A m$^2$ kg$^{-1}$]. In general, the thermoremanent magnetization (TRM) recording efficiency, $\chi_{TRM}$, is the linear scaling factor relating the external, recorded magnetic field, $\boldsymbol{B}$, to a material's TRM, $\boldsymbol{M_{TRM}}$ [A m$^{-1}$], as:

$$\boldsymbol{M_{TRM}} = \chi_{TRM}\boldsymbol{B}/\mu_0 \quad \text{(S1)}$$

As demonstrated from laboratory experiments and detailed in the Supplementary Material of (1), $\chi_{TRM}$ can be related to measurements of $M_{RS}$ through the following:

$$\chi_{TRM} = \mu_0\rho_s M_{RS}/a \quad \text{(S2)}$$

where $\rho_s$ is the density of the sample and $a$ is an experimentally-determined proportionality constant defined by the relationship $B\rho_s M_{RS}/M_{TRM} = a$ (1). For Fe-metal and FeNi-metal phases (i.e., likely candidates for Hermean crustal magnetic carriers; see Introduction), $a$ has empirically been found to be ~$3\times10^{-3}$ T (2, 3).

The aubrite achondrites are plausible analogs for Mercury crustal magnetic lithologies because they have similar elemental iron abundances (~1-10 wt% total Fe; with Fe also stored in non-ferromagnetic phases like troilite) (4), their magnetic carriers are dominated by Fe-metal and FeNi-metal grains (5, 6), and they are thought to have formed in similarly reducing conditions (7). Therefore, to define the range of $\chi_{TRM}$ for Mercury crustal material with equation (S2), we used the range of $M_{RS}$ values reported for various aubrites in (5) [see also Figure 8 of (8)], ~0.00025 – 0.016 A m$^2$ kg$^{-1}$, giving $\chi_{TRM}$ of ~0.0003 - 0.02 [SI]. We note that this upper-bound for the aubrite-$\chi_{TRM}$ range (~0.02) is the same value as was calculated from iron meteorite $M_{RS}$ measurements and scaled to the available ~1.5 wt% Fe in the Hermean crust in ref. (9).

**3D-Magnetohydrodynamic (MHD) Model**

Our 3D-MHD simulations were performed with the Block-Adaptive Tree Solarwind Roe-type Upwind Scheme (BATS-R-US) library (10, 11), utilizing the same Mercury-ancient solar wind interaction model as presented in (12, 13). We chose MHD simulations for modeling the solar wind interaction with Mercury's magnetic field given their extensive use for studying this environment [e.g, (14-16)] and noting that the ancient solar wind ion gyroradius would have been <0.01 Mercury radii ($R_M$) (see Table S2). The simulations solve the ideal and resistive MHD equations (S3)-(S6) [see (17)] to second-order accuracy until a steady-state is reached [see Fig. S8].

$$\frac{\partial\rho}{\partial t} + \nabla\cdot(\rho\boldsymbol{u}) = 0 \quad \text{(S3)}$$

$$\frac{\partial(\rho\boldsymbol{u})}{\partial \mathrm{t}} + \nabla\cdot\left(\rho\boldsymbol{u}\boldsymbol{u} + p\mathrm{I} + \frac{B^2}{2\mu_0} - \frac{\boldsymbol{B}\boldsymbol{B}}{\mu_0}\right) = \rho\boldsymbol{g} \quad \text{(S4)}$$

$$\frac{\partial\boldsymbol{B}}{\partial t} + \nabla\times\boldsymbol{E} = 0 \quad \text{(S5)}$$

$$\frac{\partial}{\partial t}\left(\frac{1}{2}\rho u^2 + \frac{p}{\gamma-1} + \frac{B^2}{2\mu_0}\right) + \nabla\cdot\left[\frac{1}{2}\rho u^2\boldsymbol{u} + \frac{\gamma p}{\gamma-1}\boldsymbol{u} + \frac{\boldsymbol{E}\times\boldsymbol{B}}{\mu_0}\right] = \rho(\boldsymbol{g}\cdot\boldsymbol{u}) \quad \text{(S6)}$$

Equations (S3)-(S6) are the conservative form of the resistive MHD equations, such that $\boldsymbol{B}$ is the magnetic field vector, $\boldsymbol{E} = -\boldsymbol{u}\times\boldsymbol{B} + \nabla\times\boldsymbol{B}/(\mu_0\sigma)$ is the electric field including the finite conductivity, $\sigma$, of the Mercury interior, $t$ is time, $\gamma$ = 5/3 is the ideal gas polytropic index, $p$ is the thermal pressure, $\boldsymbol{g}$ is the Mercury gravitational acceleration vector, $\rho$ is the plasma mass density, $\boldsymbol{u}$ is the plasma bulk velocity vector, $\mathrm{I}$ is the identity matrix, and $\mu_0$ is the magnetic permeability of free space. We note that the non-bolded variables are the scalar magnitudes of the corresponding bolded (vector) quantities. Outside Mercury's interior, the ideal MHD equations are solved (i.e., $\sigma\to\infty$), whereas inside the body, the plasma $\boldsymbol{u}$ and $\rho$ were set to zero, solving for the resistive induction-diffusion magnetic field evolution (equation S5) with a conductivity structure as described in [ (14-16), see Fig. S1 in (12)]. Equations (S3)-(S6) were solved on a spherical grid with outward logarithmically scaled cell size in the radial direction, allowing for better resolution in the planet surface and interior. The computational domain extended from radius $0.8R_M$ to $20R_M$ (outer boundary) to fully allow for the magnetospheric cavity (if applicable) to form. The grid resolution was further increased with BATS-R-US' mesh refinement capabilities for the cells within $R_M$. The inner boundary was found at the top of the highly conducting Hermean core (radius of $0.8R_M$). The radial cell size at $R$ = 0.8, 1.0, and 20 $R_M$ was 0.008, 0.011, 0.9 $R_M$, respectively, while the angular cell size at $R$ = 0.8, 1.0, and 20 $R_M$ was 0.012, 0.016, and 1.2 $R_M$. The cell resolution at the surface and in the interior was similar to that of previous BATS-R-US Mercury simulations, capturing the relevant magnetospheric and core induction physics (14-16).

The polar and azimuthal (angular) boundary conditions were periodic along the polar axis and zero meridian. At the core-mantle boundary, the boundary conditions were set to the constant dipole value and any perturbations to the magnetic field are reflected as the core can be approximated as a perfectly conducting sphere (18). The Hermean surface was not a boundary to the magnetic field, because the induction-diffusion equation transitions from the (plasma) perfect conductivity regime outside the body into the finite conductivity regime for the crust and mantle. At the surface, the inflowing solar wind was absorbed, and any outflow was inwardly reflected, as the Hermean surface is not a significant source of plasma. At the outer edge of the domain, the boundary conditions for the solar wind were set to inflow or outflow based on the direction of the solar wind velocity relative to the boundary normal vector.

To mitigate the violation of the divergence free condition of the magnetic field from numerical discretization errors in the solution of the induction equation, we employed both the eight-wave and the hyperbolic cleaning methods (10, 11). A semi-implicit time-integration scheme was used to solve the magnetic induction-diffusion equation inside the body, while outside the body, an explicit scheme was utilized with Courant number 0.8 (10).

| Number (Fig. 5) | Dynamo Mechanism | Weak Field? | Axisymmetric Field? | Offset Dipole? | Consistent with Present-Day Field and Stronger Ancient Dynamo? |
|---|---|---|---|---|---|
| 1 | solar wind feedback (19-21) | Yes | Yes | No | No |
| 2 | thermoelectric (22, 23) | Yes | No | No | No |
| 3 | iron-snow (24-26) | Yes | Yes | No | No |
| 4 | thick-shell convection (27) | Yes | Yes | No | No |
| 5 | thin-shell convection (28, 29) | Yes | No | No | No |
| 6 | convection with variable heat flux (30) | No | Yes | Yes | No |
| 7 | double diffusive convection (31) | Yes | Yes | Yes | Yes |
| 8 | convection with a stable layer and laterally variable heat flux (32) | Yes | Yes | Yes | Yes |
| 9 | convection under a thick stable layer with uniform heat flux (33) | Yes | Yes | Yes | Yes |

**Table S1: Summary of dynamo mechanisms and their consistency with a stronger ancient dynamo.** The first two columns list the dynamo mechanisms as numbered in the Discussion section and as depicted in Figure 5 ("Number"). The next three columns summarize the findings of each mechanism to reproduce the three oddities of Mercury's present-day dynamo field: weak field, axisymmetric field, and offset dipole. The last column summarizes our finding that only the double diffusive convection (31), convection with a stable layer and laterally variable heat flux (32), and convection under a thick stable layer with uniform heat flux (33) mechanisms produce all three oddities of the present-day field and represent possible evolutionary endmembers of an ancient, nearly fully convective ancient dynamo. This table is adapted from Table 1 of ref. (33).

| Run # | $B_{IMF, X}$ (nT) | $B_{IMF, Y}$ (nT) | $B_{IMF, Z}$ (nT) |
|---|---|---|---|
| 1 | 0 | 0 | 300 |
| 2 | 0 | 0 | -300 |
| 3 | 220 | -190 | 80 |
| 4 | 220 | 190 | -80 |
| 5 | 220 | -190 | -80 |
| 6 | -220 | -190 | -80 |
| 7 | -220 | 190 | 80 |
| 8 | 220 | 190 | 80 |
| 9 | -220 | 190 | -80 |
| 10 | -220 | -190 | 80 |

**Table S2**. **Solar wind parameters for 3D-MHD simulations to derive surface magnetic field timeseries.** Listed are values of components of interplanetary magnetic field for used as initial conditions for 10 MHD simulations (denoted by Run #) of interaction with both the $B_{planet}$ = 200 nT and 2000 nT dipole fields. The solar wind and IMF parameters are derived in Section 3 of (12). Here the IMF components are in the planetocentric Cartesian coordinates ($B_{IMF, X}$, $B_{IMF, Y}$, $B_{IMF, Z}$), (Figure S3). For all runs, the solar wind mass density (ρ), temperature ($T$), velocity in $X$-direction ($u_X$, Fig. S1) were fixed at ρ = 200 amu cm$^{-3}$, $T$ = $2\times10^6$ K, and $u_X$ = 1000 km s$^{-1}$, respectively.

| Layer | Depth range (km) | $\rho$ (kg m$^{-3}$) | $k$ (W m$^{-1}$ K$^{-1}$) | $C_p$ (J kg$^{-1}$ K$^{-1}$) |
|---|---|---|---|---|
| Regolith | 0-2 | 2350 | 0.3 | 1000 |
| Crust | 2-35 | 2900 | 2.0 | 1000 |
| Mantle | >35 | 3380 | 4.0 | 1142 |

**Table S3**. **Parameters for cooling models.** Listed are depth range, material density ($\rho$), thermal conductivity ($k$), and specific heat capacity ($C_p$) used for the cooling simulations of the *Regolith*, *Crust*, and *Mantle* layers. Regolith density is estimated assuming a 19% porosity, consistent with the 18-20% range assumed in previous studies [e.g., (34)]. Together with an assumed equilibrium surface temperature of 440 K and heat flux of 25 mW m$^{-2}$, these approximately reproduce the expected equilibrium temperature profile for Mercury's interior at 3.0 Ga (35).

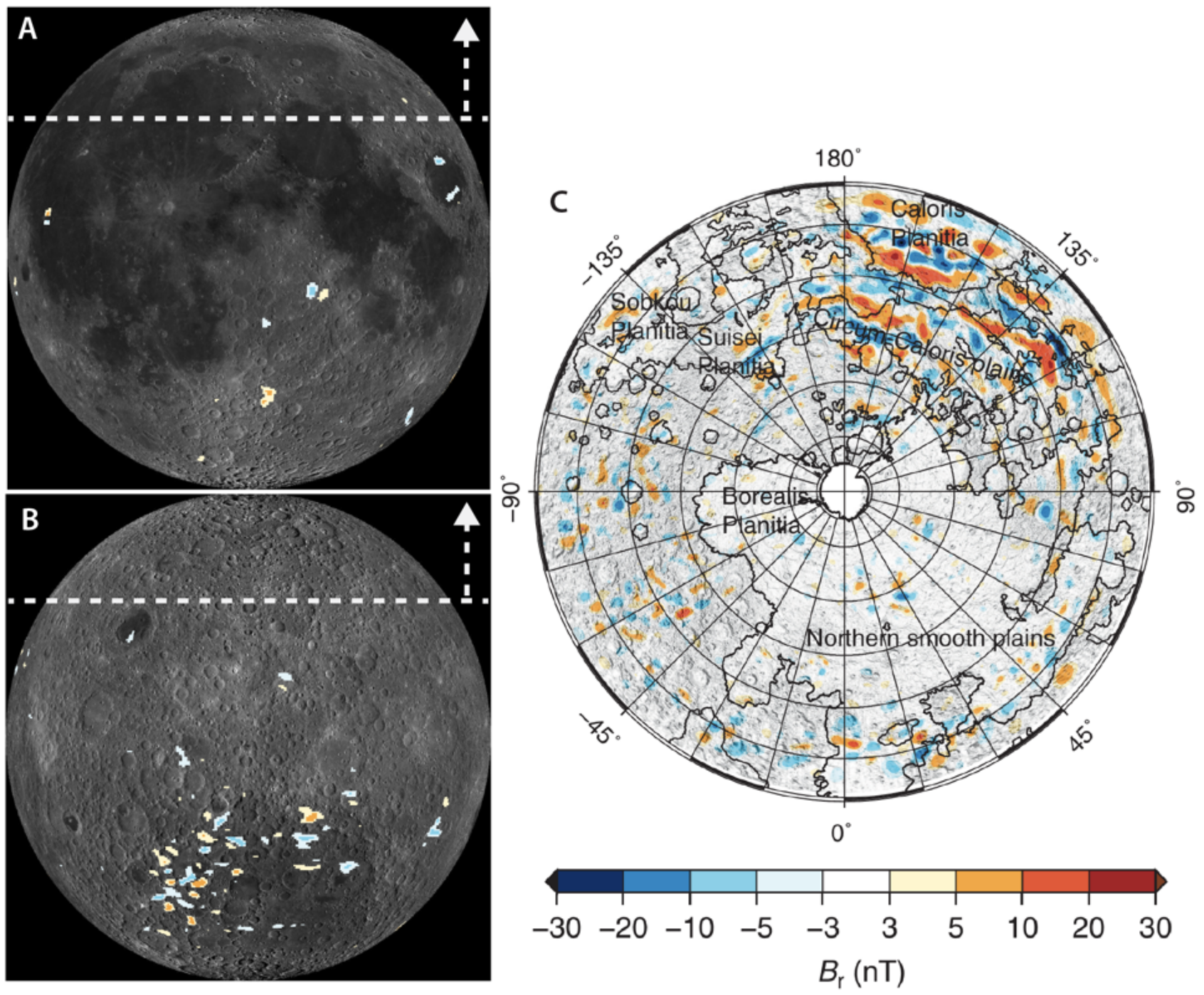

**Figure S1. Orbital crustal magnetic field maps of the Moon and Mercury.** Shown are the crustal magnetic field radial component ($B_r$ at 30-km altitude) for the Moon [centered (**A**) nearside and (**B**) farside, adapted for *Lunar QuickMap* (36) from ref. (37)] and ($B_r$ at 20-km altitude) for Mercury [**C**: 32°N-90°N, adapted from Fig. 5.19 in ref. (38)], plotted on the same color scale. The white dashed lines and arrows represent the lunar 32°N-90°N latitude region, the same latitude range as seen in the Mercury map. While there are multiple mechanisms invoked to explain the setting of the lunar anomalies [e.g., (17, 39)], one such mechanism attributes these structures to impactor-delivered metals (e.g., Fe-metal) cooling in the presence of an ancient dynamo field [see (40) and refs. therein]. The presence of exogenic Fe-metal is invoked partially due to the poor magnetic recording efficiencies ($\chi_{TRM}$) of native lunar materials, which can be >2 orders of magnitude less efficient than the magnetic recorders in the meteorite collection (1). The delivery of these impactor materials for the lunar magnetization process is plausible given the sparse nature of these anomalies, which are also commonly co-located with impact structures (37, 40, 41). Conversely, Mercury's measured crustal fields are, compared to their lunar counterparts, relatively strong, widespread (covering much greater surface area), and heterogeneously distributed with respect to geologic structure (e.g., volcanic plains and impact craters). While it is possible that some of Mercury's surface anomalies could contain Fe-metal wt% higher than the global average [~1.5 wt% (42-44)], it is unlikely that this mechanism alone is able to explain all strong anomalies. This motivates our modeling choice of $\chi_{TRM}$ scaled to the average surface Fe-metal wt%.

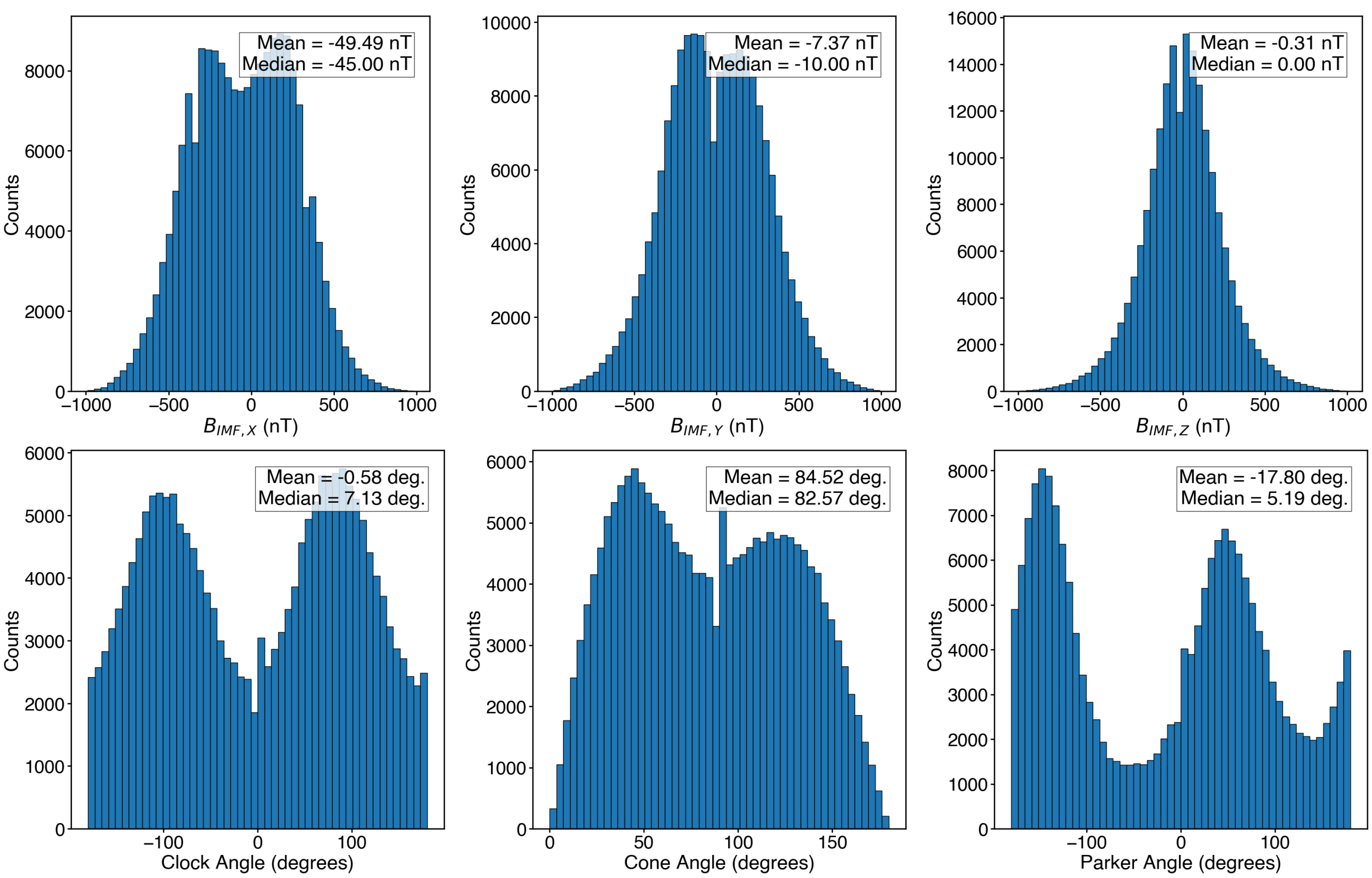

**Figure S2. Distribution of derived ancient and modern interplanetary magnetic field (IMF) components and orientation angles at Mercury's orbit ~3.9-3.7 Ga ago.** Hourly counts of the modern IMF (top row) Cartesian planetocentric coordinate system (Figure S3) components ($B_{IMF,\,X}$, $B_{IMF,\,Y}$, and $B_{IMF,\,Z}$) and (bottom row) IMF orientation angles measured by the ACE spacecraft at the Lagrange point 1 over a 20-year period, scaled to the inferred strength (~500 nT) of IMF at the orbit of Mercury ~3.9-3.7 Ga ago (called "scaled ACE"). The ACE spacecraft is taken to be the center of this "planetocentric" Cartesian coordinate system (Figure S3). The magnetic field components are expected to all center around zero magnitude as the IMF has directional variability on much shorter timescales than the total time of the set. The Parker ("spiral") angle specifies the relative sense of the radial (*X*-direction, Sun to Mercury direction) and azimuthal (*Y*-direction, Mercury orbital direction) magnetic field components, where an angle of -135° or 45° means that these two components are similar in magnitude, as is seen here for the modern IMF at Earth. Our calculations of the average Parker angle for the ancient IMF at Mercury yielded a value of ~40°**,** giving further confidence that the "scaled ACE" dataset is a good approximation for the IMF ~3.9-3.7 Ga ago at Mercury. The clock angle gives a sense of the relative dominance between the azimuthal (*Y*-direction) and vertical (*Z*-direction) components of the IMF. The cone angle gives a sense of the dominance of the radial (*X*-direction) magnitude of the IMF.

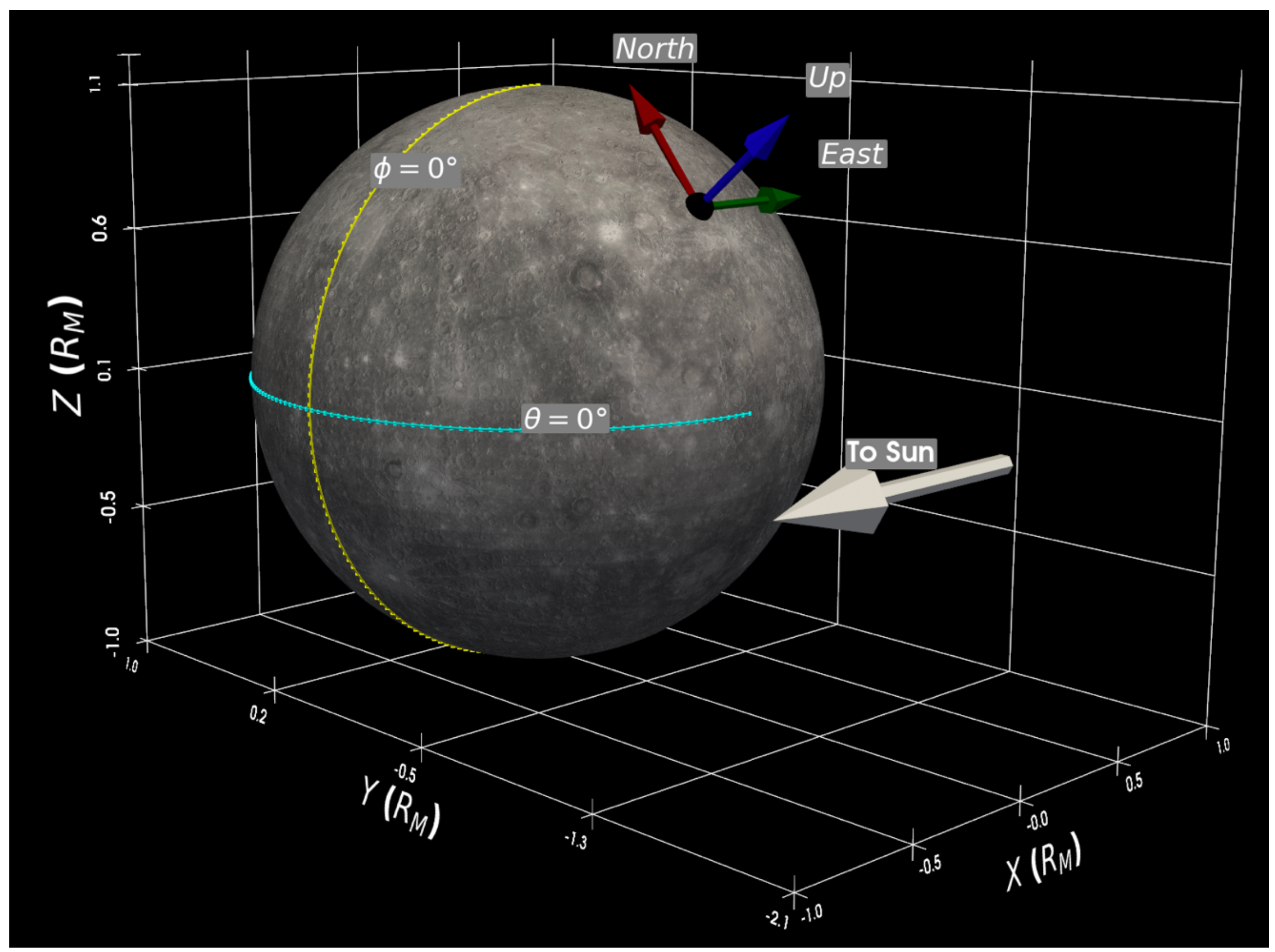

**Figure S3. Geometry for Mercury MHD and thermal cooling and magnetization simulations.** Planetocentric, Mercury-Sun-orbital Cartesian (*X*, *Y*, and *Z*) coordinate system and local tangent plane coordinate system [(green) *East*, (red) *North*, and (blue) *Up*] shown with the Hermean surface in units of the planet radius, $R_M$. The grey arrow represents the direction towards the Sun, such that the solar wind flows in the +*X*-direction. The gold curve shows the zero-longitude line ($\phi$ = 0 with $\phi$ increasing eastward) and the light blue curve shows equator ($\theta$ = 0), which define the Sun-fixed (planetocentric) latitude-longitude coordinate system used to map the 3D magnetic field solution to the timeseries.

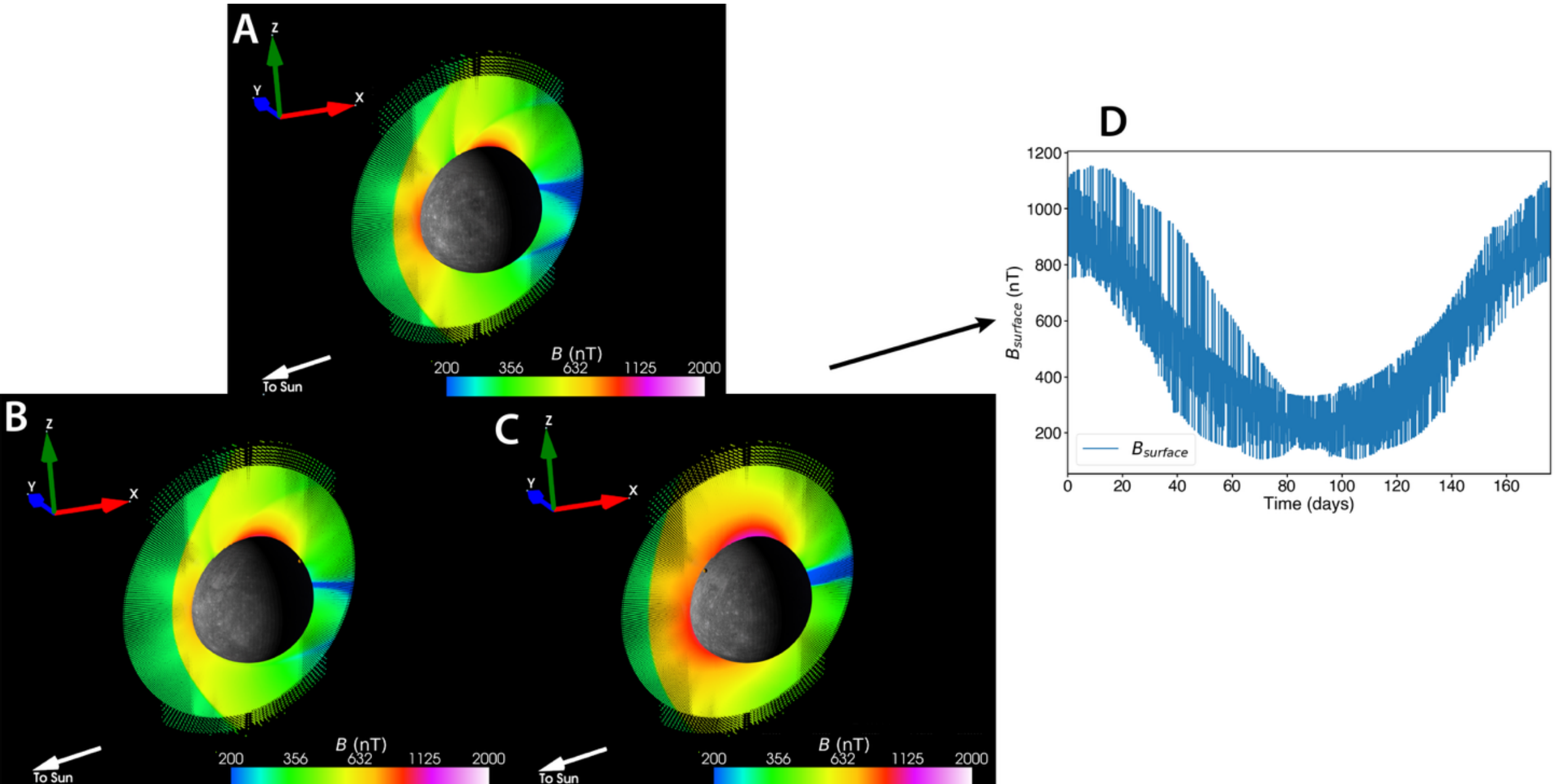

**Figure S4. MHD simulations of Mercury's magnetic field environment used to derive surface field timeseries.** Two-dimensional slices of three 3D-MHD simulation snapshots (**A**, **B**, and **C**) of magnetic field used to derive (**D**) surface magnetic field ($B_{surface}$) time series incorporated into thermal and cooling magnetization model (Movie S1). In each of these simulations, the planetary dipole field was set to $B_{planet}$ = 200 nT at the surface equator. This time series was derived from a set of 10 MHD simulations for this dipole field. This process was repeated for the $B_{planet}$ = 2,000 nT dipole field. The time series was generated from sampling the magnetic field solutions of the 10 simulations, mapping longitude to time (see Materials and Methods).

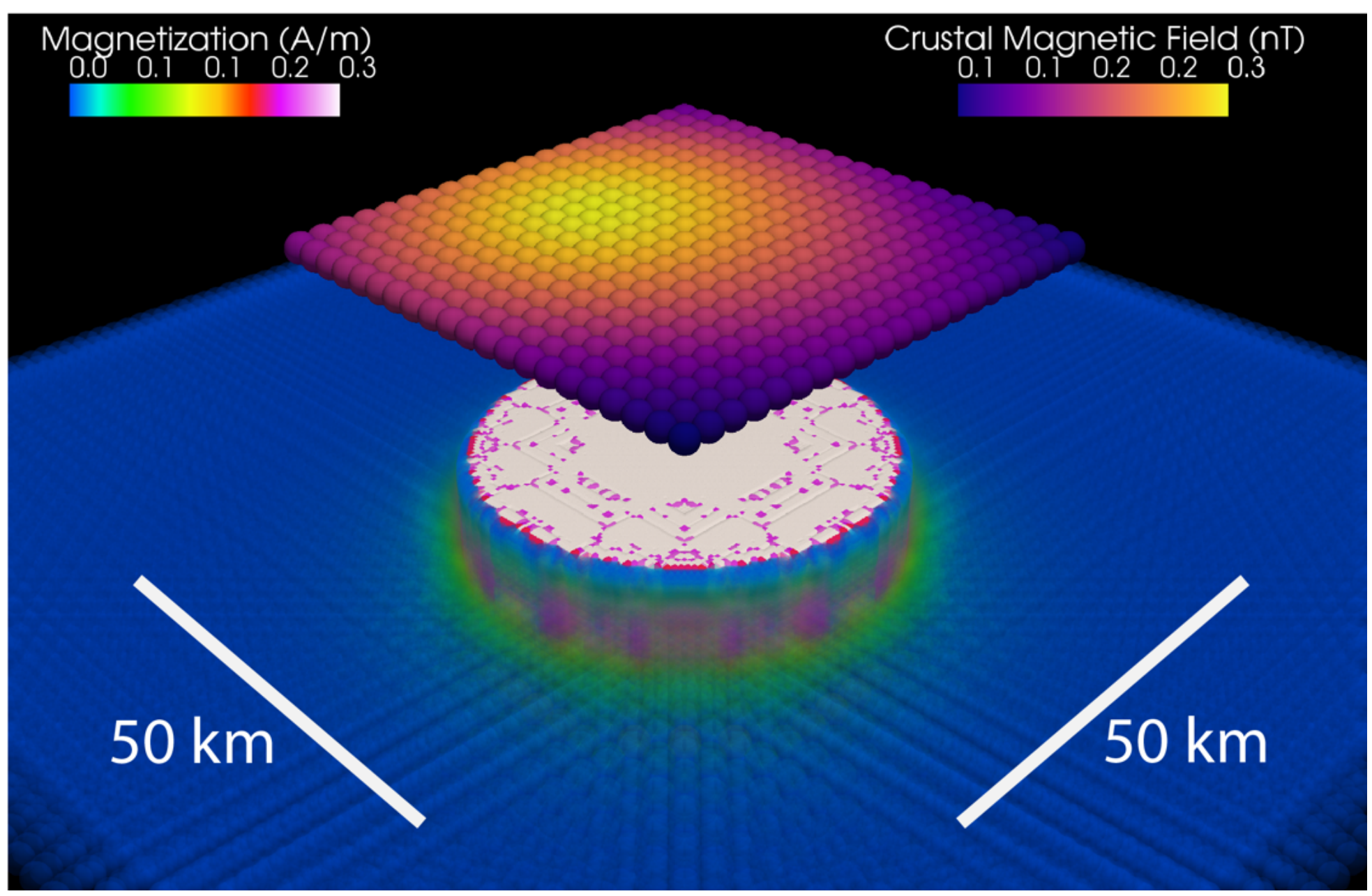

**Figure S5. Magnetized volume and crustal remanent field at 30-km altitude.** 3D map of DSK50 volume magnetization (A/m) magnetized by the magnetic field timeseries with the $B_{planet}$ = 200 nT present-day field and its generated crustal magnetic field (nT) at 30-km altitude. This map shows the spatial scale of the recorded magnetization at the completion of the cooling process of the DSK50 volume. This magnetized volume results in a maximum crustal generated magnetic field of ~0.3 nT at 30-km altitude, well below the strong MESSENGER-level measured signals.

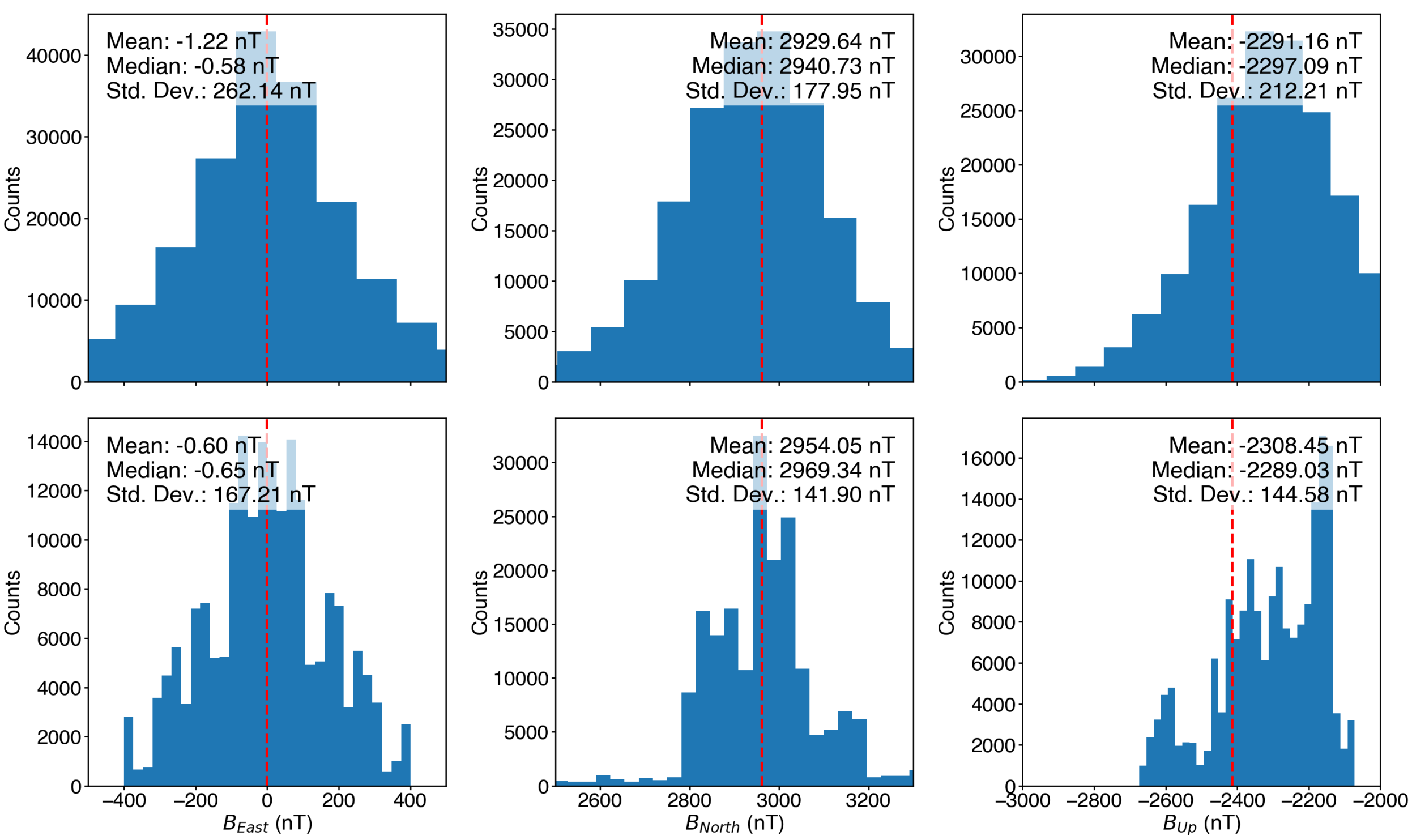

**Figure S6. Normal and direct sampling distributions of derived surface magnetic field timeseries for the 2,000 nT surface dipole scenario.** Counts of the local tangent plane magnetic field components (left, middle, and right: $B_{East}$, $B_{North}$, and $B_{Up}$; Figure S3) for the (top) normal distribution and (bottom) direct sampling, both defined by results at 40° latitude from the ten 3D-MHD simulations run with the 2,000 nT dipole field (see Materials and Methods for further discussion). The red vertical lines represent the fixed dipole field magnitudes at these latitudes, which are constant over all longitudes due to the axisymmetry of the Hermean dynamo field. Both sampling methods result in similar distribution shapes, medians, means, and standard deviations (listed in the legend of each subplot), and as such, result in similar magnetizations and generated crustal fields, holding $\chi_{TRM}$ and magnetized volume constant.

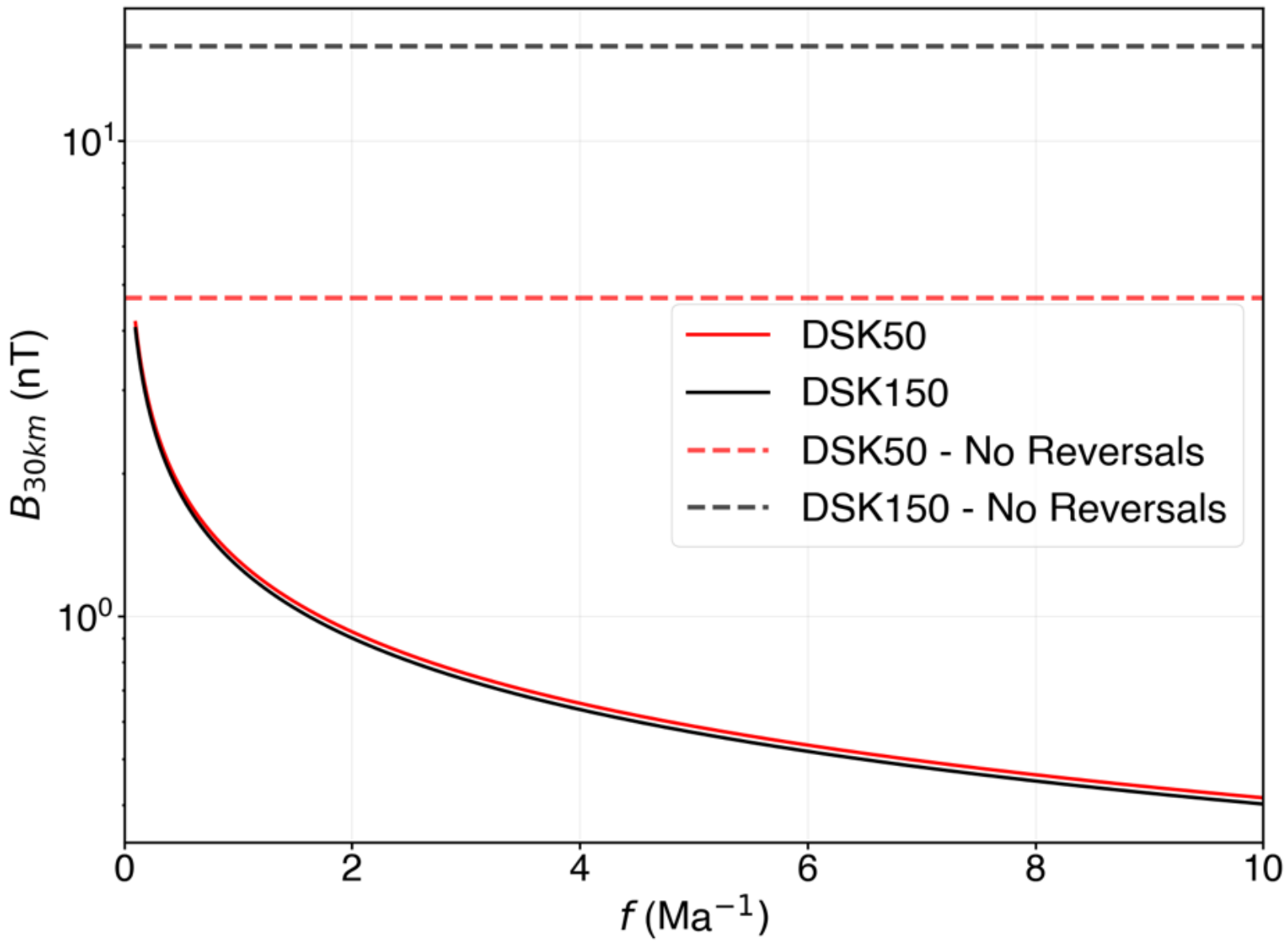

**Figure S7. Dependance of the remanent field of magnetized cylinders on the reversal frequency of the dynamo magnetizing source.** Relationships between the DSK50 (red) and DSK150 (black) generated crustal fields at 30-km altitude and the reversal frequency ($f$) of the dynamo source field. The dashed lines show the nominal crustal field strength for the DSK50 (red) and DSK150 (black) volumes sourced by a uniform field of 3,700 nT (field strength of dipole at 40° N for 2,000 nT equatorial surface dipole). For $f$ = 0.1 Ma$^{-1}$, the DSK50 crustal field is minimally diminished (<1.2×) while the DSK150 crustal field is lessened by >4×, both compared to the uniform source field remanent field (dashed lines). For $f$ = 10 Ma$^{-1}$, the DSK50 resultant crustal field is diminished by >20× while the DSK150 crustal field is lessened by >80×, both compared to the uniform source field remanent field (dashed lines). All crustal field values are computed with recording efficiency $\chi_{TRM}$ = 0.02.

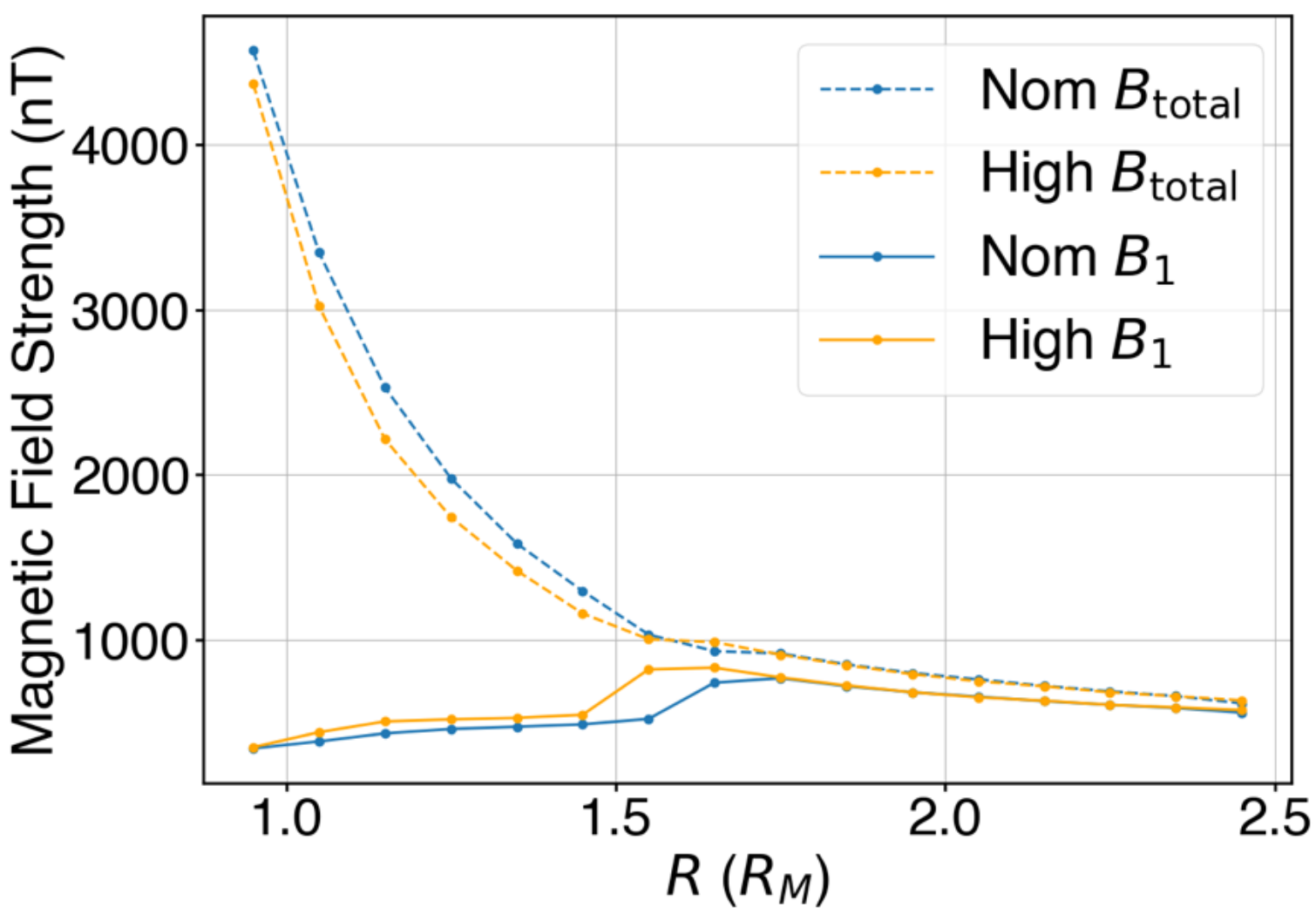

**Figure S8. Magnetic field solution from two grid resolutions showing solution convergence.** Line plot depicting 3D-MHD solution radial coordinate ($R$ in units of Mercury radius, $R_M$) versus magnetic field strength for (blue curves) nominal grid resolution ("Nom") simulations used to derive the surface magnetic field timeseries in this study and for (orange curves) a 2× higher (in each dimension) grid resolution ("High") simulations. The dashed and solid curves show the solution total magnetic field ($B_{total}$) and the perturbation magnetic field solution ($B_1$), respectively, which reflects the non-dipole field contributions to $B_{total}$ [see section 2.3.2 of (10) for description of "split magnetic field"]. The 3D-MHD simulations used the parameters as described in "Run #6" in Table S2 with the 2,000 nT dipole field as measured at the equatorial surface. These line plots are subsequently calculated by averaging the magnetic field strength over all cells with equivalent radial coordinates at latitude of 40°N, where $R$ = 1.0 on this plot represents the average field in the generated magnetic field timeseries (see Fig. S6). The nominal grid resolution simulation ("Nom") has Mercury surface cells of radial size ~30-km and both spherical angular sizes of ~40-km. The higher grid resolution simulation ("High") has Mercury surface cells of radial size ~16-km and both spherical angular sizes of ~20-km. Importantly, this plot shows that increasing the nominal grid resolution by 2× in each spherical coordinate only changes the magnetic field solution by <10%, giving confidence that the numerical solution has converged and that further resolution would not change our main result.

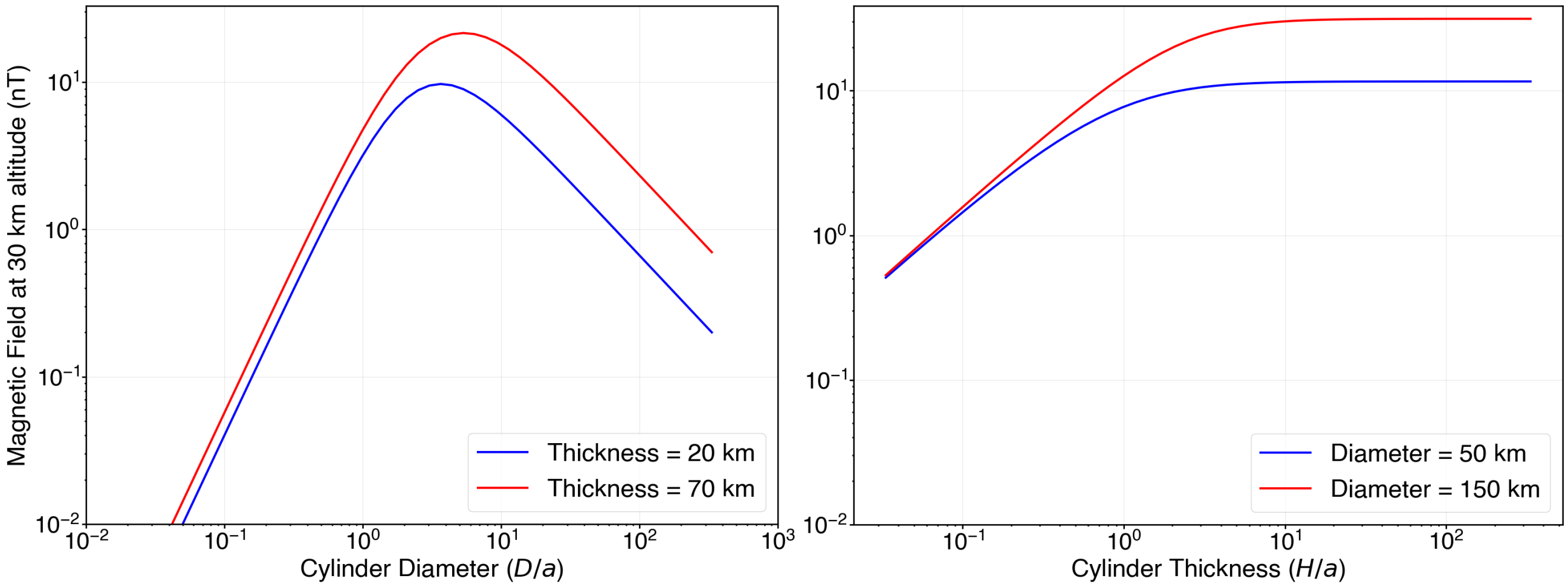

**Figure S9. Dependence of the remanent field of magnetized cylinder on its diameter and thickness.** Relationships between (left) uniformly magnetized cylinder diameter, *D*, and (right) thickness, *H*, for generating a 10 nT crustal field (*B*) at 30-km altitude, *a*. These magnetized volumes are set by a constant 10,000 nT field directed along the cylindrical long axis with recording efficiency $\chi_{TRM}$ = 0.02. The calculations (left) for varying *D* hold *H* constant with values (blue line) 20-km and (red line) 70-km. Similarly, the calculations (right) for varying *H* hold *D* constant with values (blue line) 50-km and (red line) 150-km. Both plots have the varying quantity on the horizontal axis normalized to the spacecraft altitude, *a*. These plots show the limiting relationships between *D*, *H*, and *B*: (*D*, *H* << *a*), (*D*, *H* ~ 2*a*), and (*D*, *H* >> *a*). For *D, H* << *a*, *B* approaches zero as these scale lengths approach zero. For *D, H* ~ 2*a*, *B* is maximized as the length scale approaches the physically-resolvable surface resolution (taken to be the spacecraft altitude). For *D* >> *a*, *B* approaches zero as the cylinder diameter approaches infinity (45). Conversely, *B* plateaus when *H* >> *a* as the volume is increasing further away from the measurement point *a*. This asymptotic behavior occurs as magnetized volume is being added at increasing distances which has negligible effect on *B* (measured at *a*) which scales with distance cubed.

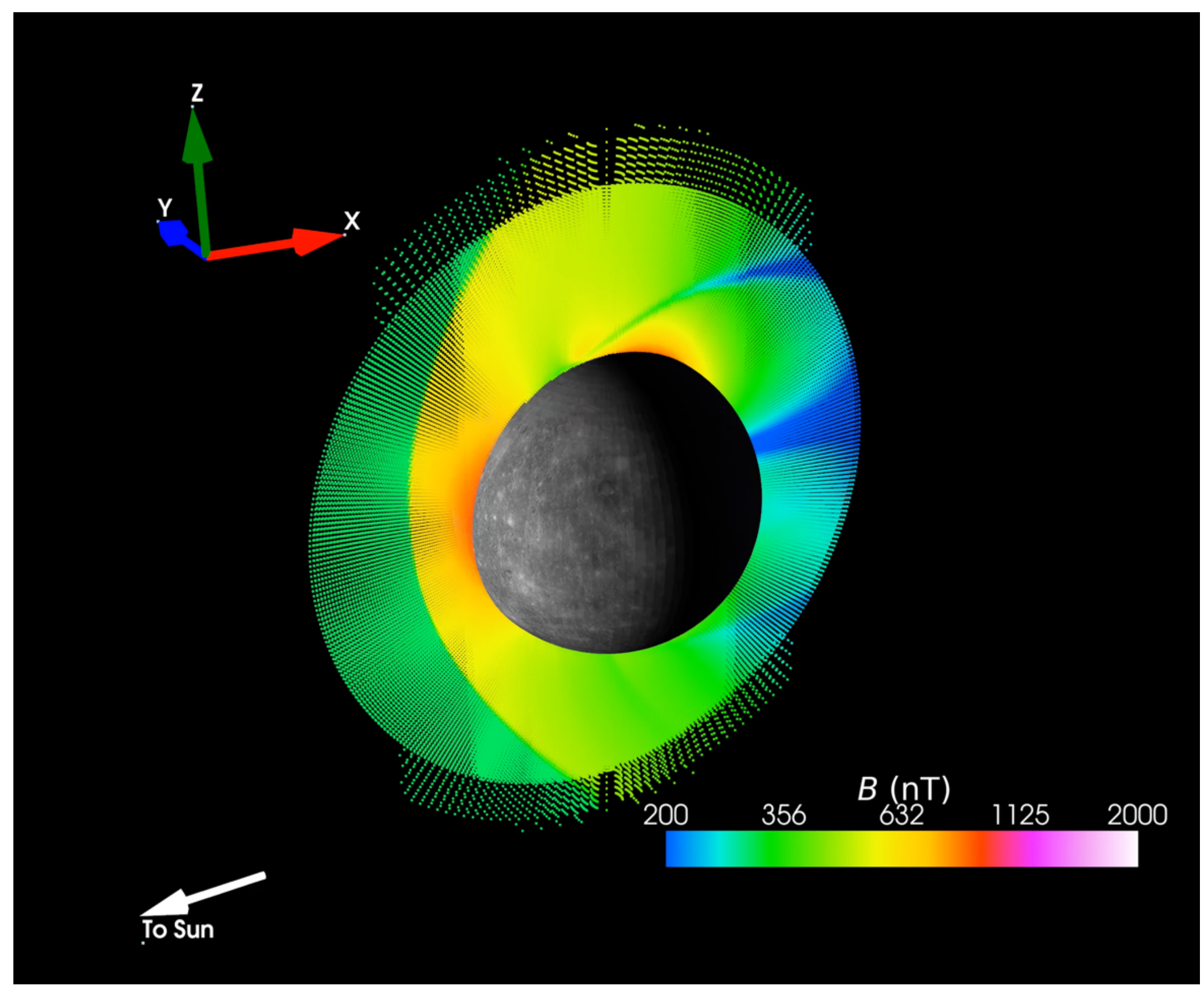

**Movie S1: Evolution of the simulated magnetic field environment over a Hermean day.** 3D-MHD visualization of 2D slices of magnetic field (*B*) from the 10 solar wind conditions (Table S2) interacting with the modern-dynamo field (~200 nT at equatorial surface). The orange “dot” on the surface represents the 40° N latitude. The planet rotating on its spin-axis under the changing solar wind depicts how the surface magnetic field timeseries are generated for a given latitude (i.e., at the orange dot) by sampling the conditions at different longitudes (see Materials and Methods, Figures 6, S4, and S6).